\documentclass[10pt,journal,compsoc]{IEEEtran}
\usepackage{multirow}
\usepackage{hhline}

\ifCLASSOPTIONcompsoc
  \usepackage[nocompress]{cite}
\else
  \usepackage{cite}
\fi
\ifCLASSINFOpdf
   \usepackage[pdftex]{graphicx}
   \graphicspath{{../pdf/}{../jpg/}}
   \DeclareGraphicsExtensions{.pdf,.jpg,.png}
\else
   \usepackage[dvips]{graphicx}
   \graphicspath{{../eps/}}
   \DeclareGraphicsExtensions{.eps}
\fi
\usepackage{amsmath}
\usepackage{amsfonts}
\DeclareMathOperator*{\argmax}{arg\,max}

\usepackage{amsthm}
\theoremstyle{definition}
\newtheorem{definition}{Definition}[section]

\usepackage{algorithmic}

\usepackage{array}

\ifCLASSOPTIONcompsoc
  \usepackage[caption=false,font=footnotesize,labelfont=sf,textfont=sf]{subfig}
\else
  \usepackage[caption=false,font=footnotesize]{subfig}
\fi
\usepackage{url}

\begin{document}
%
\title{An Affective Situation Labeling System from \\Psychological Behaviors in Emotion Recognition}
%
%

%
%

\author{Byung~Hyung~Kim
        and~Sungho~Jo,~\IEEEmembership{Member,~IEEE}
\IEEEcompsocitemizethanks{\IEEEcompsocthanksitem B. Kim and S. Jo are with the School of Computing, KAIST, Republic of Korea. S. Jo is the corresponding author.\protect\\
E-mail: \{bhyung, shjo\}@kaist.ac.kr}
}

\IEEEtitleabstractindextext{%
\begin{abstract}
This paper presents a computational framework for providing affective labels to real-life situations, called A-Situ. We first define an affective situation, as a specific arrangement of affective entities relevant to emotion elicitation in a situation. Then, the affective situation is represented as a set of labels in the valence–arousal emotion space. Based on physiological behaviors in response to a situation, the proposed framework quantifies the expected emotion evoked by the interaction with a stimulus event. The accumulated result in a spatiotemporal situation is represented as a polynomial curve called the affective curve, which bridges the semantic gap between cognitive and affective perception in real-world situations. We show the efficacy of the curve for reliable emotion labeling in real-world experiments, respectively concerning 1) a comparison between the results from our system and existing explicit assessments for measuring emotion, 2) physiological distinctiveness in emotional states, and 3) physiological characteristics correlated to continuous labels. The efficiency of affective curves to discriminate emotional states is evaluated through subject-dependent classification performance using bicoherence features to represent discrete affective states in the valence–arousal space. Furthermore, electroencephalography-based statistical analysis revealed the physiological correlates of the affective curves.
\end{abstract}

}

\maketitle

\IEEEdisplaynontitleabstractindextext

%
\IEEEpeerreviewmaketitle

\IEEEraisesectionheading{\section{Introduction}\label{sec:introduction}}

%
%
%
%
\IEEEPARstart{E}{motion} with supervised training datasets has received much attention in recent years, because it facilitates the understanding of emotional interactions between humans and computers by measuring emotional states such as joy, excitement, and fear. However, obtaining a massive amount of well-labeled data is usually very expensive and time-consuming. Although there have been advances in the annotation of emotional states in various environments, most cases depend on the participant's self-assessment~\cite{trull2013ambulatory,shiffman2000real,shiffman2008ecological}. Apart from some existing issues with validity and corroboration (e.g., participants may not answer with exactly how they are feeling, but instead give responses similar to those they expect others would likely provide)~\cite{bethel2007survey}, this kind of reporting can only gather immediate human affective output in numerical form, providing only a limited understanding of complex emotional conditions and affective dynamics in daily life. Hence, it is critical to provide an automatic method for labeling human emotions elicited in real-life situations

However, quantifying emotional responses based on the understanding of emotional interactions in real-world situations is challenging. It requires a cognitive understanding of the real-world objects that humans interact with and a determination of the expected affective level of the humans’ emotions based on the interaction. In response to this challenge, we start by defining the term ``affective situation,'' as a specific arrangement of affective entities in a spatiotemporal domain. Affective entities can be any of the real-world objects that people encounter and interact with in a place at a given time. Next, we present a computational framework to model and represent affective situations for labeling of real-life situations, called A-Situ. To model affective situations, the system derives pairs of emotion labels in the valence–arousal space from low-level features extracted from a psychological behavior sequence in a target situation.

Our model is mainly intended to estimate emotional adaptability to a situation in order to label emotional states underlying 1) affective response, 2) approach and withdrawal motivation, and 3) self-contentment, based on the extracted features in a sequence. The proposed framework represents an affective situation as a polynomial curve called the ``affective curve,'' which is fitted to a set of points over the valence–arousal emotion space. Furthermore, we aim to model and represent affective situations in real-world environments. To gather such environmental information, we design a wearable device that can be comfortably worn to allow users to act freely in everyday situations. consisting of a frontal camera, an accelerometer, and small physiological sensors. We use the data collected from our device to learn and represent affective situations and to provide proper affective labels to support learning of physiological changes in emotion recognition. Furthermore, modeling affective situations allows us to understand life content or material in human interaction, and representing these situations can determine the level of a person’s expected feeling based on the interaction.

The distinct contributions of the proposed system, called A-Situ, as against existing systems are as follows:

\begin{itemize}
	\item Affective Situation Representation: We introduce a polynomial curve called the ``affective curve,'' which is a set of cumulative points on the valence–arousal emotional space over time in a situation and represents affective dynamics in real-world environments.
	\item Affective Situation Modeling: Given a psychological behavior sequence in a given situation, we detect the expected feeling and track its changes. To model changes in the situation, we present three components: motivation, motion, and contentment. They reflect emotional responses to a situation's underlying low-level features. 
	\item Physiological experiments to validate the effects of affective labels produced by A-Situ as ground truth: We evaluate the proposed system over a long time series of life-logging data, covering multiple days in real-world scenarios. The evaluation involves investigating and analyzing the characteristics of brain signals related to different affective labels. Electroencephalography (EEG)–based statistical analysis reveals that physiological responses correlate to continuous affective labels.
\end{itemize}
The rest of this paper is organized as follows: In Section 2, we provide a theoretical background and overview of previous studies in emotion recognition related to affective labeling. Section 3 presents our A-Situ system, with the following subsections: 1) affective situation learning, 2) affective situation representation. In Sections 4 and 5, we evaluate the performance of A-Situ using a real-world dataset collected using our wearable device and explore how brain activity is correlated with emotional changes annotated by our system. In Section 6, we show some interesting cases that involve analyzing physiological characteristics to demonstrate the effectiveness of our system. We conclude this article with perspectives on future work.

\section{Background}
Providing labels with emotional tagging enhances multidisciplinary areas, processing different types of multimedia data such as images, videos, and texts. It quantifies affective responses to stimuli underlying affect dimensions in two ways: explicit and implicit tagging.

\begin{figure}[t]
\centering
\subfloat[]{\includegraphics[width=1.5in]{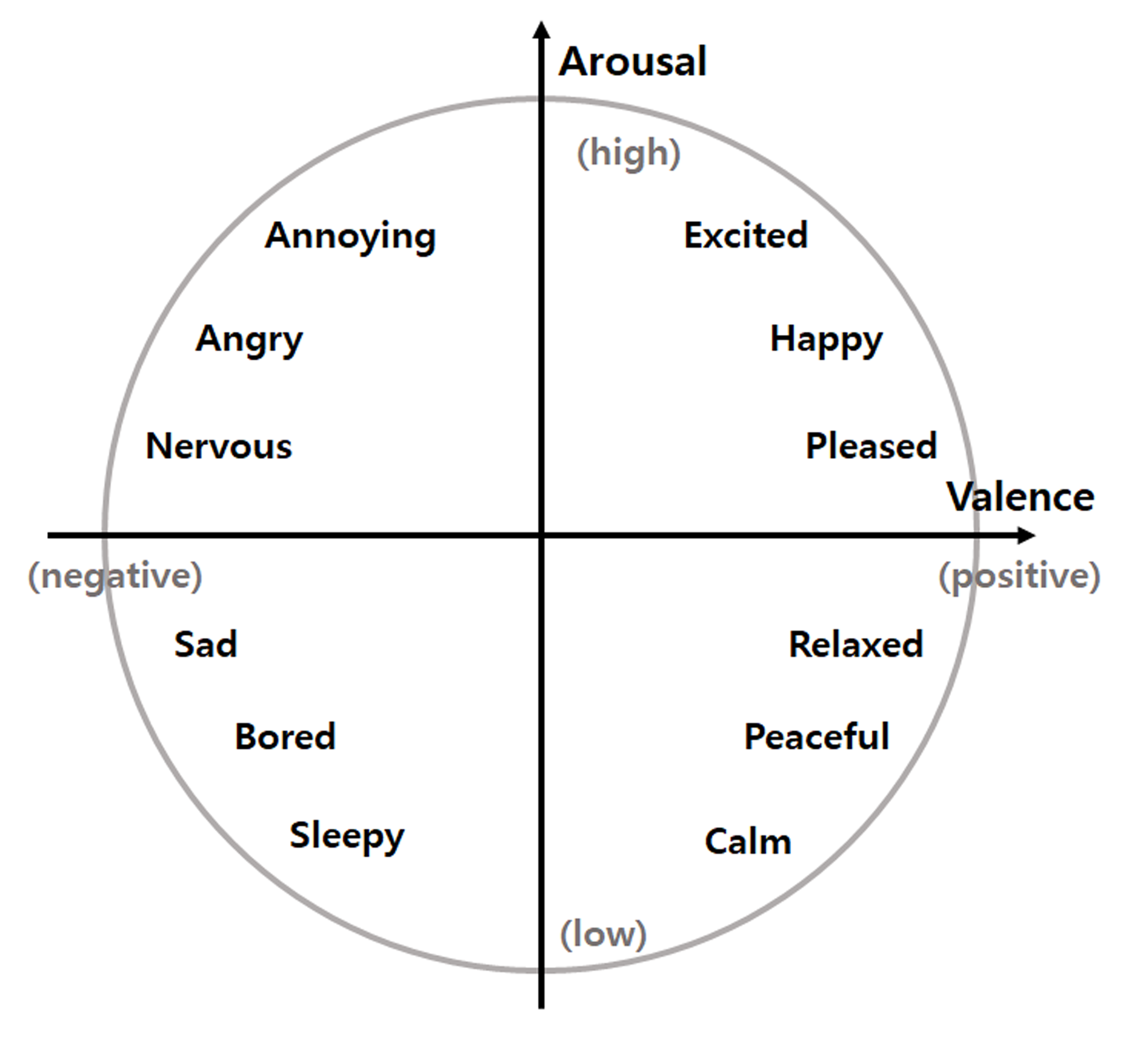}
\label{fig:VAspace}}
\subfloat[]{\includegraphics[width=1.5in]{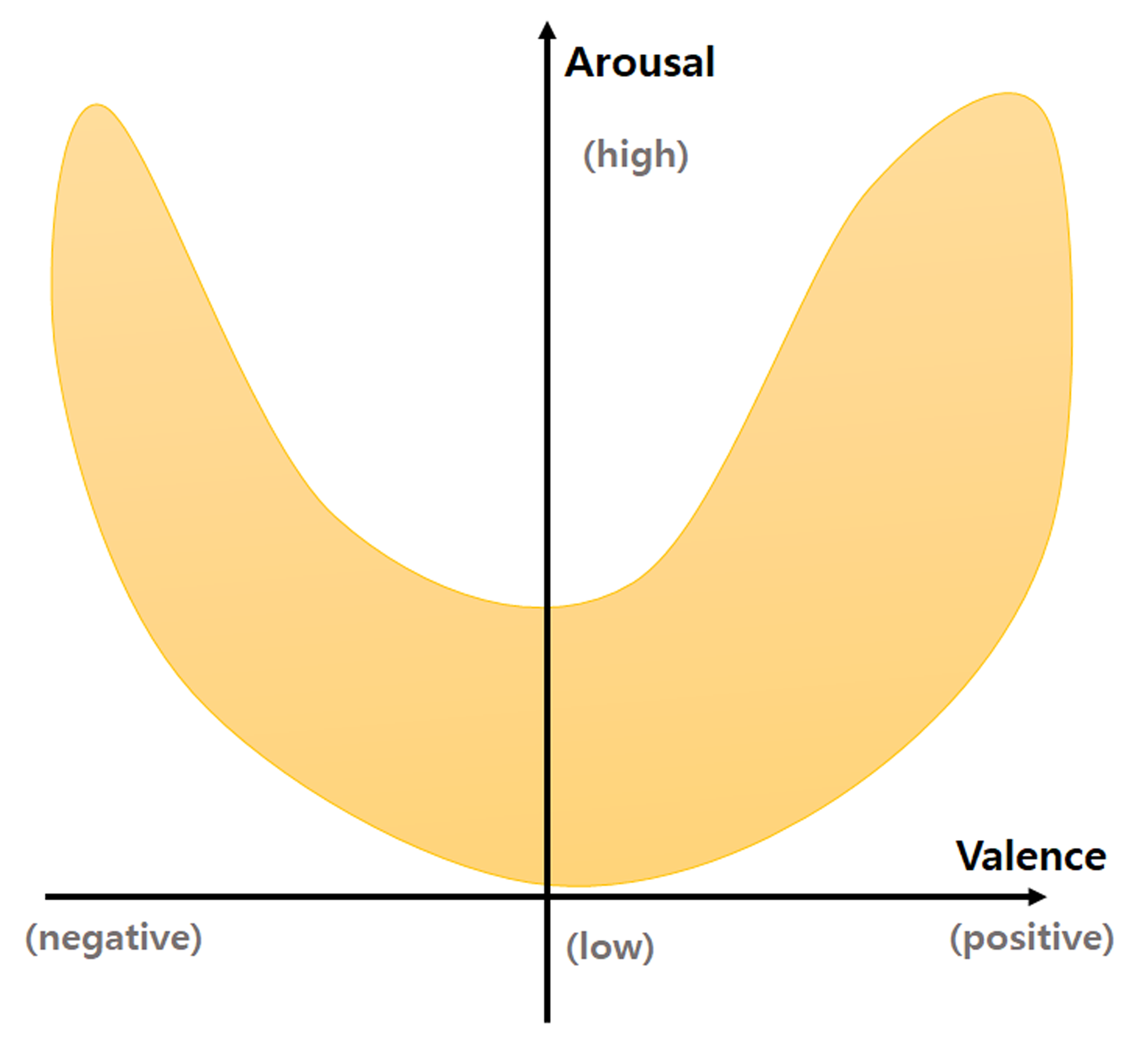}
\label{fig:paracurveVAspace}}
\caption{(a) Distribution of emotions in valence-arousal (V-A) space (b) Parabolic shape of the V-A emotion space}
\label{fig:AffectDimension}
\end{figure}

According to Bradley~\cite{bradley1994emotional} and Russel and Mehrabian~\cite{russell1977evidence}, human emotion can be conceptualized in three major dimensions of connotative meaning: valence (V), arousal (A), and dominance (D). Valence refers to the type of emotion and characterizes emotional states or responses ranging from unpleasant or negative feelings to pleasant, happy, or positive feelings. Arousal is the intensity of emotion and characterizes emotional states or responses ranging from sleepiness or boredom to frantic excitement. Dominance distinguishes emotional states having similar valence and arousal, ranging from ``no control'' to ``full control''. For instance, the emotions of grief and rage have similar valance and arousal values but different dominance values. The entire scope of human emotions can be represented as a set of points in the three-dimensional (3D) VAC coordinate space. Conversely, each basic emotion can be represented as a bipolar entity~\cite{russell1980circumplex}, characterizing all emotions by valence and arousal, and different emotional labels can be plotted at various positions on this two-dimensional V–A plane (see \figurename~\ref{fig:VAspace}).

Although several studies aim to collect a wide range of emotions using audio-visual content~\cite{kim2007neural,britton2006facial}, recent studies have found that affective responses mapped onto the emotional coordinate system are roughly parabolic (see \figurename~\ref{fig:paracurveVAspace})~\cite{kuppens2013relation, lithari2010females}. For example, Dietz and Lang~\cite{dietz1999affective} used the parabolic surface to assign temperament, mood, and emotion to define the personality of an affective agent.

\subsection{Affect Labeling: Explicit and Implicit Methods}
The explicit approach provides labels by asking users to report their feeling in response to given events or stimuli. For instance, the International Affective Picture System (IAPS) has been a popular dataset; in it, an explicit self-reporting tool such as the SAM has been used to acquire affective labels~\cite{lang1997international}. The SAM is a picture-based assessment technique used to measure emotional response to a wide variety of stimuli associated with valence from positive to negative, arousal from high to low, and dominance from low to high. Dynamic assessments, such as ambulatory assessment~\cite{trull2013ambulatory} and ecological momentary assessment~\cite{shiffman2008ecological}, allow the opportunity to assess contextual information about a behavior, and serve as real-time self-report methods to measure behavior and experiences in people’s daily lives. The assessments collect data from various devices, such as smart phones and mobile physiological devices. For instance, for the assessment of emotions and cognitions associated with eating habits, participants may be asked to answer questions on a smart phone each time it beeps and before and after all meals and snacks. Affective labels obtained from explicit self-reporting tools have been considered ground-truth data for emotional states~\cite{sharma2017continuous} and used to build reliable emotion recognition systems~\cite{kim2018deep}. At the same time, a major drawback of the explicit approach to labeling human emotions is the intrusiveness of the reporting procedure. Furthermore, obtaining a massive amount of hand-labeled data is very expensive and time-consuming.

Conversely, the implicit affective labeling approach is unobtrusive, as labeling is obtained by exposing users to stimuli and recording their responses. In emotion recognition work, visual and motion features have been important elements for tagging emotions in different types of multimedia data, such as images and videos. Joho et al.~\cite{joho2011looking} used facial change characteristics to label human emotions.  Simmons et al.~\cite{simons1999emotion} studied object motion as a visual feature in response to human affect and showed that increasing the motion intensity could also lead to increased levels of emotional arousal. Zhang et al.~\cite{zhang2010affective} developed a method to characterize arousal using motion intensity and shot change rate in video clips. Hanjalic et al.~\cite{hanjalic2005affective} used motion activity to determine arousal levels and represented continuous change of arousal as a curve. However, implicit methods like these have limits as far as a cognitive understanding of the real-world objects that humans interact with, since they have perceived emotions based on the scene as ``understanding.''

\subsection{Psychological Behaviors}
An alternative to the above implicit approaches is to extract emotional features of psychological behaviors and associate them with emotional states. In this paper, we focus on developing psychological components in response to stimuli. Approach–avoidance theory describes action tendencies in response to emotion evoked by a stimulus event. The main proposition of the theory is that approach tendencies emerge toward positive stimuli and avoidance tendencies for negative stimuli. Krieglmeyer and Deutsch~\cite{krieglmeyer2010comparing} conducted experiments to compare measures of approach–avoidance behaviors for sensitivity and criterion-validity. The results showed that a manikin task outperformed joystick tasks in this regard due to the means of distance change, such as (the manikin) running towards the object instead of (the joystick) moving it. Many studies have proposed methods to label emotional difference based on psychological behaviors. For example, arm movements such as flexion and extension have been investigated to reveal positive and negative interactions between emotional stimuli and responses to approach and avoidance behaviors~\cite{laham2015meta}. Seibt et al.~\cite{seibt2008movement} used a joystick to determine whether positive and negative stimuli facilitate approach and withdrawal behaviors, respectively. Participants were instructed to control the joystick by either pulling it to increase the size of the stimuli or pushing it to decrease the size. Seibt et al.'s metric could discriminate between approach and avoidance behaviors in teenagers reacting to a positive or negative stimulus. However, the studies cited here are restricted to controlled experimental settings, require the use of specific equipment, and use limited-perception tasks in which participants are not interacting in real time with the system. In contrast, our system aims to label emotions by detecting the expected feeling and tracking its changes from a psychological behavior sequence in real-world situations.

\subsection{Physiological Sensors in Emotion Recognition}
Physiological measurement has been a key to understanding emotions. Several studies on emotion detection have advanced significantly in many ways over the past few decades~\cite{sander2005systems}. EEG measurement refers to the recording of the brain's electrical activity with multiple electrodes placed on the scalp. Its very high temporal resolution is valuable to real-world applications despite its low spatial resolution on the scalp~\cite{lovato2014meta}. Moreover, mobility techniques of non-invasive EEG have extended their usage to the field of brain-computer interfaces (BCIs), external devices that communicate with the user’s brain~\cite{wolpaw2012brain}.  Peripheral physiological signals such as skin conductance, heart rate, and breathing rate have been also carried out in emotion assessment~\cite{subramanian2016ascertain}. In these measurements, distinct or peaked changes of physiological signals in the autonomic nervous system (ANS) elicited by specific emotional states at a single instantaneous time have been considered as candidates. Due to the simplicity, they have been used to develop wearable biosensors in clinical applications such as detecting mental stress in daily life~\cite{zhang2015troika}. However, this approach is limited and cannot be used to fully describe emotion elicitation mechanisms due to their complex nature and multidimensional phenomena. In our work, EEG is the most suitable choice among available physiological measurements since it measures the brain dynamics that control thoughts, feelings, and behaviors.
\section{Affective Situation Labeling System}
\begin{figure*}[h]
    \centering
    \includegraphics[width=1.8\columnwidth]{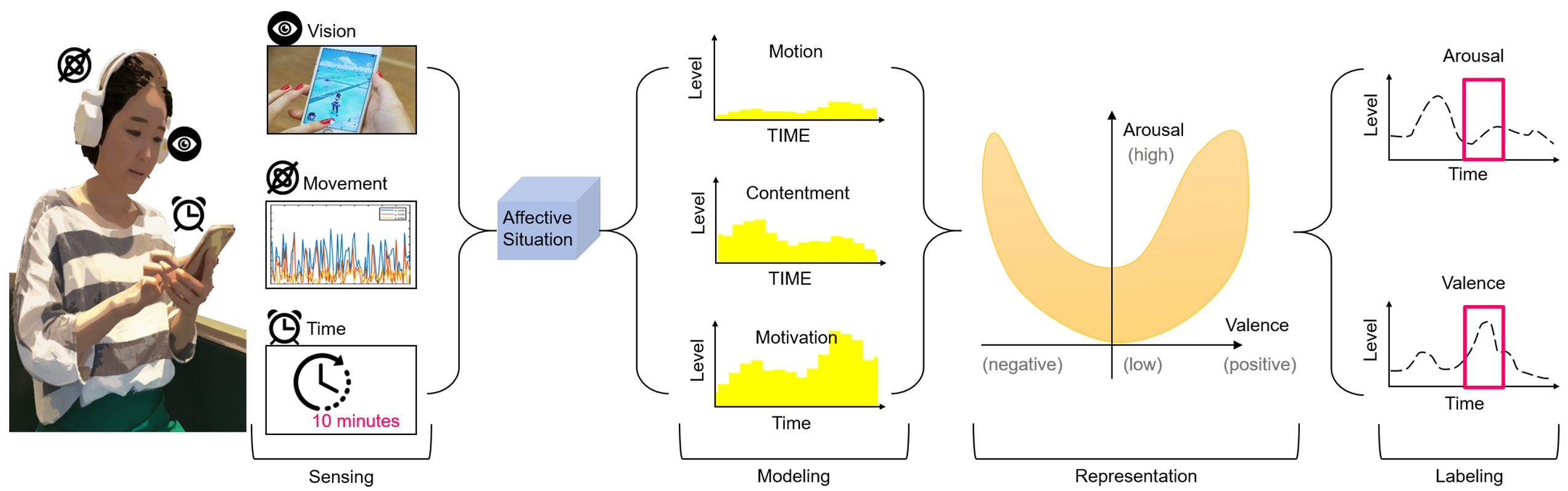}
    \caption{Overview of the proposed A-Situ system. For every timestamp, our system recognizes the expected feeling based on a person's behaviors in a situation. A set of expected feelings in an affective situation is represented by a curve called affective curve.}
    \label{fig:overview}
\end{figure*}
A-Situ defines an affective situation in order to represent and model it as a set of points in the valence-arousal emotion space.

\begin{definition}{Affective Situation:}
An affective situation $\mathcal{S}^i_t$ is a specific arrangement of affective entities relevant to emotion elicitation in situation $i$ at time $t \in T_i$ 
\small
\begin{equation}
    \mathcal{S}^i_t =(\mathcal{M}^i_t, \mathcal{E}^i_t, T_i),
\end{equation} where $\mathcal{M}^i_t$ is an egocentric image sequence, $\mathcal{E}^i_t$ is an accelerometer sequence, $T_i$ is the length of situation $i$.
\end{definition}

\figurename\ref{fig:overview} shows the entire framework of A-Situ. The system provides affective labeling from an affective situation in real-world scenarios. To quantify the feeling evoked in a situation, A-Situ focuses on learning and representing an affective situation. At each time $t$, our system takes an egocentric image $\mathcal{M}^i_t$ and uses auxiliary accelerometer data $ \mathcal{E}^i_t$ sequences as inputs, outputting a set of two emotional points $\mathcal{L}^i_t$ over valence-arousal space. The learned points are represented as a polynomial curve called affective curve.
\small
\begin{equation}
    \hat{\mathcal{L}}_{1:t}=\argmax_{(\mathcal{V},\mathcal{A}) \in \mathcal{L}}p(\mathcal{L}_{1:t}|\mathcal{S}_{1:t})
    \label{eq:problemStatement}
\end{equation}

\subsection{Affective Situation Learning}
In a given situation, we can observe several affective expressions. As described in Section 2, the following factors can be used to model these emotional phenomena in terms of arousal and valence:
\begin{itemize}
    \item Motion: The influence of object motion on human emotional response has revealed that an increase in motion intensity causes an increase in arousal. 
    \item Motivation: Some theories regard affective valence to be tightly coupled with motivational direction, such that positive affect is associated with approach motivation and negative affect is associated with avoidance motivation
    \item Contentment: Attitudes toward discrete emotions predict emotional situation selection. For instance, more positive attributes toward ``excited'' are more likely to express interest in adapting ``excited''-evoking stimuli with self-contentment. 
\end{itemize}
Based on the three factors, A-Situ produces valence $\mathcal{V}$ and arousal $\mathcal{A}$ values, imposing spatial constraints on valence-arousal space, based on the following criteria.  
\begin{itemize}
  \item Comparability: This ensures that the values of arousal, valence, and the resulting affect curve obtained in different situations for similar types of emotional behavior are comparable. This criterion naturally imposes normalization and scaling requirements when computing time curves.
  \item Compatibility: This ensures that the shape of the affect curve reflects the situation at a particular given time in the valence–arousal emotion space. When the situation ends, the appearance of the curve becomes a roughly parabolic contour of the 2D emotion space.
  \item Smoothness: This describes the degree of emotional retention of preceding frames in the current frame. It ensures that the affective ratio of the content related to eliciting human emotions does not change abruptly between consecutive frames of a situation.
\end{itemize}

The proposed system uses general functions $\mathcal{A}(\mathcal{S})$ and $\mathcal{V}(\mathcal{S})$ for arousal and valence in an affective situation $\mathcal{S}$. The two functions have the appropriate form of functions to integrate the three components: motion, motivation, and contentment components as given above. 

\subsubsection{Motion Component of Emotional Responses}
To calculate the motion component $m(\mathcal{S}^i_t)$, A-Situ estimates the motion of objects in situation $i$ at time $t$. The system first uses optical flow estimation to characterize and quantify the motion of affective objects between adjacent frames; then, the average magnitude of all estimated motion vectors formulates motion activity 
\begin{small}
\begin{equation}
\label{eq:motion1}
\bar{m}(\mathcal{S}^i_t)=\frac{1}{B|\vec{v}_{max}|}(\sum^B_{k=1}{|\vec{v_k}(t)|}),
\end{equation}
\end{small}where $\vec{v_k}(t)$ is the motion vector $k$ and $B$ is the number of motion vectors at time $t$. To suppress motion artifacts, we used accelerometer data $\mathcal{E}^i_t$ in the motion component $m(\mathcal{S}^i_t)$.
\begin{small}
\begin{equation}
\label{eq:motion2}
m(\mathcal{S}^i_t) = (1 - G( \mathcal{E}^i_t ))\cdot \bar{m}(t),
\end{equation}
\end{small}where $G(\cdot)$ is the Gaussian smoothed results normalized between 0 and 1. Note that $1 - G( \mathcal{E}^i_t )$ implies that an increase in motion artifacts causes a decrease in arousal, because motion artifacts are not actual factors of motion activities.

\subsubsection{Motivation Component of Emotional Approach-Withdrawal Behaviors}
\begin{figure*}[t]
    \centering
    \includegraphics[width=1.8\columnwidth]{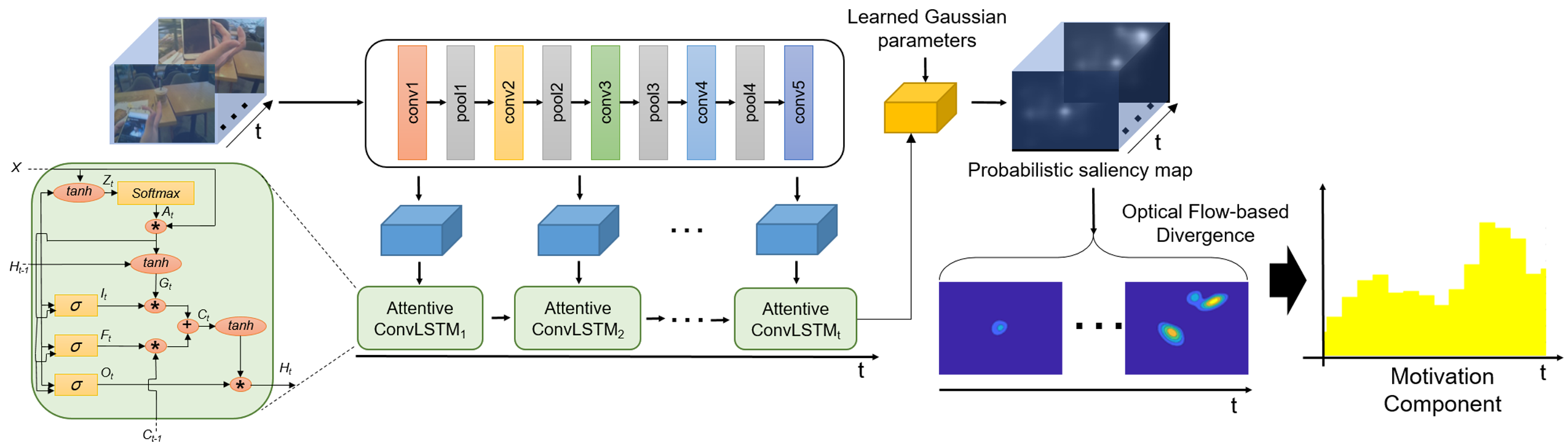}
    \caption{Overview of the motivation component. From an image at each video time $t$, visual attention is detected in the human cortex by the saliency maps. Under the area covered by the binary saliency map, the emotional approach-withdrawal behaviors associated with the attentive object are calculated using optical flow around the object.}
    \label{fig:overviewMotivation}
\end{figure*}
\begin{figure}[t]
    \centering
    \includegraphics[width=0.8\columnwidth]{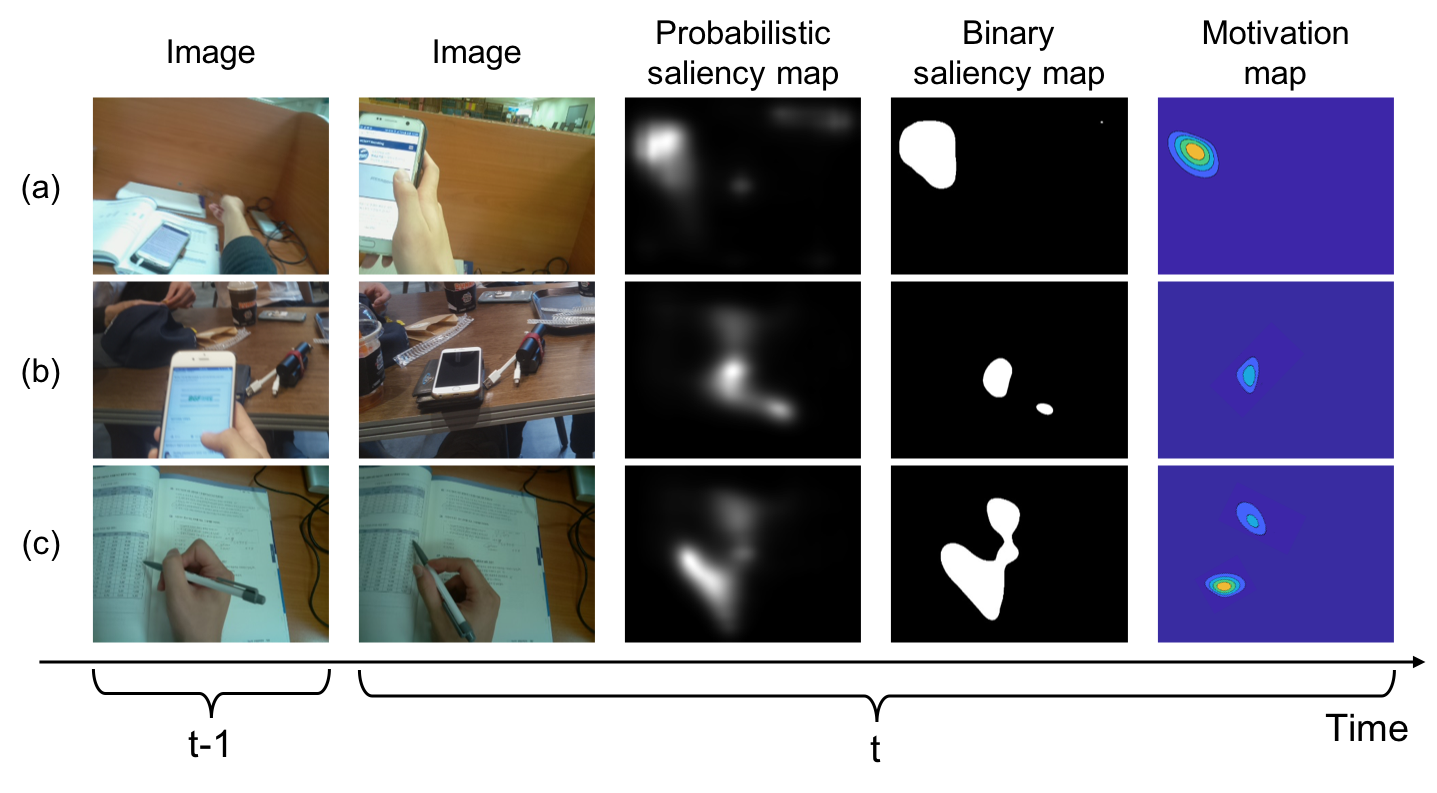}
    \caption{Motivation component examples. (a) Raising a hand while interacting with a mobile phone has only positive effects on the value for the component, while (b) laying down has the effect of decreasing values in the situation, and (c) moving a hand slightly and rotating a pen has a complex motivational effect on the component.}
    \label{fig:MotivationComponentExamples}
\end{figure}
The motivation component is derived in two stages. It aims to compute emotional saliency within visually attentive areas. We first detect the participant's intention regarding visual attention. Predicting the location of visual attention maintained at a certain fixation point can be done with saliency prediction or detection. To obtain the most salient region in an image frame, we used the saliency-attentive (SA) model, as in \cite{cornia2016predicting}, in which human eye fixations during a scene were predicted by building a convolutional long short-term memory (ConvLSTM) with a set of features computed by dilated convolutional networks (DCN) and multiple learned gaze priors as a salient object detector.

In an affective situation, ConvLSTMs take visual features extracted from images and refine them in the prior learned module. More specifically, they compute an attention map by convolving the previous hidden state and the input, producing the output as a normalized saliency spatial map through the softmax operator in the output layer. Given the final saliency map, which is a probability map with values within [0, 1], we generate a binary saliency map with a threshold $t_h$. Then, the white area in the binary saliency map becomes the prime fixation area to which the participant applies visual attention. \figurename~\ref{fig:MotivationComponentExamples} shows saliency predictions for sample images using the output of a ConvLSTM module at different timesteps as an input to the rest of the model. Within the area of saliency prediction, we compute emotional saliency after the second stages.  

As the second stage, we learn the emotional approach–withdrawal behaviors associated with a saliency object in the prime fixation area detected by the first stage. More specifically, we compute divergence and rotation using optical flow around the attentive  object at each video frame $t$. An approach to a single object can be identified by zooming in on the object, and this has the same effect on the divergence of flow vectors surrounding the center point of the object~\cite{li2009recognition}. Inversely, avoidance of a single object associated with withdrawal behaviors can be estimated by the convergence, which is tantamount to zooming out from the object.

To compute the divergence, we first compute the flow using multi-scale block-based matching between adjacent frames. Then, the flow is standardized as six primitive optical flow patterns~\cite{li2009recognition}: 1) rotation around a vertical axis; 2) rotation around a horizontal axis; 3) approach toward an object; 4) rotation around the optical axis of the image plane; and 5) and 6) complex hyperbolic flows. Given the motion vector field, the velocity $e_p$ at the pixel $p(x,y)$ can be represented as

\begin{small}
\begin{equation}
\label{eq:motivation1}
e_p = e_{p_0} + \bar{\chi}(p-p_0),
\end{equation}
\end{small}where $e_{p_0} = (u_1, u_2)^T$ is the velocity at pixel $p_0$ and $\bar{\chi}$ is the matrix defined as
\begin{small}
\begin{equation}
\label{eq:motivation2}
\bar{\chi}=\begin{bmatrix} \chi_1 & \chi_3 \\ \chi_2 & \chi_4 \end{bmatrix}.
\end{equation}
\end{small}
Then $\bar{\chi}$ can be decomposed to 
\begin{small}
\begin{equation}
\label{eq:motivation3}
\bar{\chi} = \frac{1}{2}(d_1D_1 + d_2D_2 + h_1H_1 + h_2H_2),
\end{equation}
\end{small}where the four matrices are defined as follows: 
\begin{small}
\begin{equation}
\label{eq:motivation4}
D_1=\begin{bmatrix} 1 & 0 \\ 0 & 1\end{bmatrix}, D_2=\begin{bmatrix} 0 & -1 \\ 1 & 0\end{bmatrix},\\
H_1=\begin{bmatrix} 1 & 0 \\ 0 & -1\end{bmatrix}, H_2=\begin{bmatrix} 0 & 1 \\ 1 & 0\end{bmatrix},
\end{equation}
\end{small}
where $d_1 = \chi_1 + \chi_4$ and $d_2=\chi_2 + \chi_3$ refer to divergent and rotating optical flows, respectively. $h_1 = \chi_1 - \chi_4$ and $h_2 = \chi_2 + \chi_3$ refer to different types of hyperbolic optical flows. The velocity of the motion vector field can be approximately characterized by six parameters: $u_1, u_2, d_1, d_2, h_1, h_2$. Given an optical flow field of the attentive object, we estimate the parameter vector using (\ref{eq:motivation3}) and the least squared error method. The parameter $u_1$ is associated with right and left rotations, $u_2$ is associated with heading up and down, $d_1$ is associated with approaching the object, and the last three parameters indicate combined motion.

Using the six parameters, we compute the motivation component $o(\mathcal{S}^i_t)$ at time $t$ as follows:
\begin{small}
\begin{equation}
\label{eq:motivation5}
o(\mathcal{S}^i_t) = \text{exp}(\sum_{x=0}^X\sum_{y=0}^Y({d_1}_{(x,y)} / ( {d_2}_{(x,y)} + {h_1}_{(x,y)} + {h_2}_{(x,y)}))),
\end{equation}
\end{small}where $X$ and $Y$ denote the width and height of the optical flow field of the attentive object. The motivation component $o(\mathcal{S}^i_t)$ in (\ref{eq:motivation5}) increases when approaching an object; otherwise, it remains near zero. For instance, as shown in \figurename~\ref{fig:MotivationComponentExamples}a, raising a hand while interacting with a mobile phone has only positive effects on the values for the component, but laying down decreases values in the situation. \figurename~\ref{fig:MotivationComponentExamples}b shows minimal values reflecting non-emotional behaviors. 

\subsubsection{Contentment Component of an Affective Situation}
We used time-varying situation lengths to reveal a connection between a user's emotion and his/her intent to adapt to a situation, reflecting self-contentment. We model the emotional contentment of the situation by deriving the function $l(\mathcal{S}^i_t)$ at time $t$ as follows:
\begin{small}
\begin{equation}
\label{eq:adaptation}
l(\mathcal{S}^i_t) = \lambda_1 \log( \delta(t) - \lambda_2 ) + \lambda_3,
\end{equation}
\end{small}where $\lambda_1, \lambda_2$, and $\lambda_3$ determine the shape of the function $l(\mathcal{S}^i_t)$. The function has logistic growth until the maximum length $T_i$ while staying in situation $i$. 

\subsection{Arousal and Valence Model}
To model arousal, the function $A(\mathcal{S}^i_t)$ uses the weighted averages to integrate the contribution of the motion $m(\mathcal{S}^i_t)$ and contentment $l(\mathcal{S}^i_t)$ along with an image sequence in an affective situation at time $t$. The function is convolved with a sufficiently long smoothing window to merge neighboring local maxima of the components through a moving average filter; the result is normalized to a range of 0 to 1. 
\begin{small}
\begin{equation}
\label{eq:VAmodel1}
A(\mathcal{S}^i_t) = \alpha_1\frac{m(\mathcal{S}^i_t)}{m_{max}} + \alpha_2 \frac{l(\mathcal{S}^i_t)}{l_{max}},
\end{equation}
\end{small}where $\alpha_w$ are the coefficients for weighting the two functions with $\sum_{w=1}^2 \alpha_w = 1$. 

The Compatibility criterion requires that the affect curve generated by combining the arousal and valence time should cover an area in the valence-arousal coordinate system that has a parabolic shape resembling the 2D emotion space. Clearly, this criterion requires the values of arousal and absolute values of valence to be related; thus, in general, the range of arousal values determines the range of absolute valence values~\cite{kuppens2013relation}. We, therefore, start the development of the valence model by defining the function $r(\mathcal{S}^i_t)$ that captures this value range dependence considering the value of arousal $A(\mathcal{S}^i_t)$ at the current time $t$
\begin{small}
\begin{equation}
\label{eq:VAmodel2}
r(\mathcal{S}^i_t) = sign(l(\mathcal{S}^i_t))A(\mathcal{S}^i_t),
\end{equation}
\end{small}
\begin{small}
\begin{equation}
\label{eq:VAmodel3}
V(\mathcal{S}^i_t) = \nu_1 \frac{r(\mathcal{S}^i_t)}{r_{max}} + \nu_2 \frac{o(\mathcal{S}^i_t)}{o_{max}},
\end{equation}
\end{small}where $r(\mathcal{S}^i_t)$ implies that the negativity of the expected feeling mainly is determined by the amount of emotional contentment in a situation. If a subject wants to avoid the situation, $l(\mathcal{S}^i_t)$ would become a small value and the expected feeling would tend to be negative. Based on this function $\mathcal{A}(\mathcal{S}^i_t)$, the valence value $\mathcal{V}(\mathcal{S}^i_t)$ is determined by the motivation component $o(\mathcal{S}^i_t)$. The function $\mathcal{V}(\mathcal{S}^i_t)$ is smoothed with the same moving average filter as the function $\mathcal{A}(\mathcal{S}^i_t)$. Note that $\nu_i$ are the weighted averages of $r(\mathcal{S}^i_t)$ and $\mathcal{V}(\mathcal{S}^i_t)$, respectively. 

\subsection{Affective Situation Representation}
We represent affective situations over 2D emotion space from valence and arousal values learned by the above calculation. We used the set of two emotional values to fit a Gaussian process regression (GPR) model, which is a nonparametric kernel-based probabilistic model that uses a linear basis function and the exact fitting method to estimate the parameters of the GPR model. This results in the production of the affective curve as a representation of an emotional trace along a situation, as perceived by a human. 

\section{Real-World Experiment}
To evaluate the performance of our system for labeling emotion, we conducted real-world experiments on university life. An wearable device was designed and distritued to participants to gather frontal images, EEG signals, and accelerometer signals in their daily life. From this information, the Affective Situation Dataset $\mathbb{S}$ is developed, consisting of the gathered data and a subset $\mathbb{S}_{\zeta} \subset \mathbb{S}$, which has labels rated by the SAM. The dataset $\mathbb{S}_{\zeta}$ was used to evaluate the proposed system, which is compared to other state-of-the-art methods in the following section. 
\begin{figure*}[t]
    \centering
    \includegraphics[width=2\columnwidth]{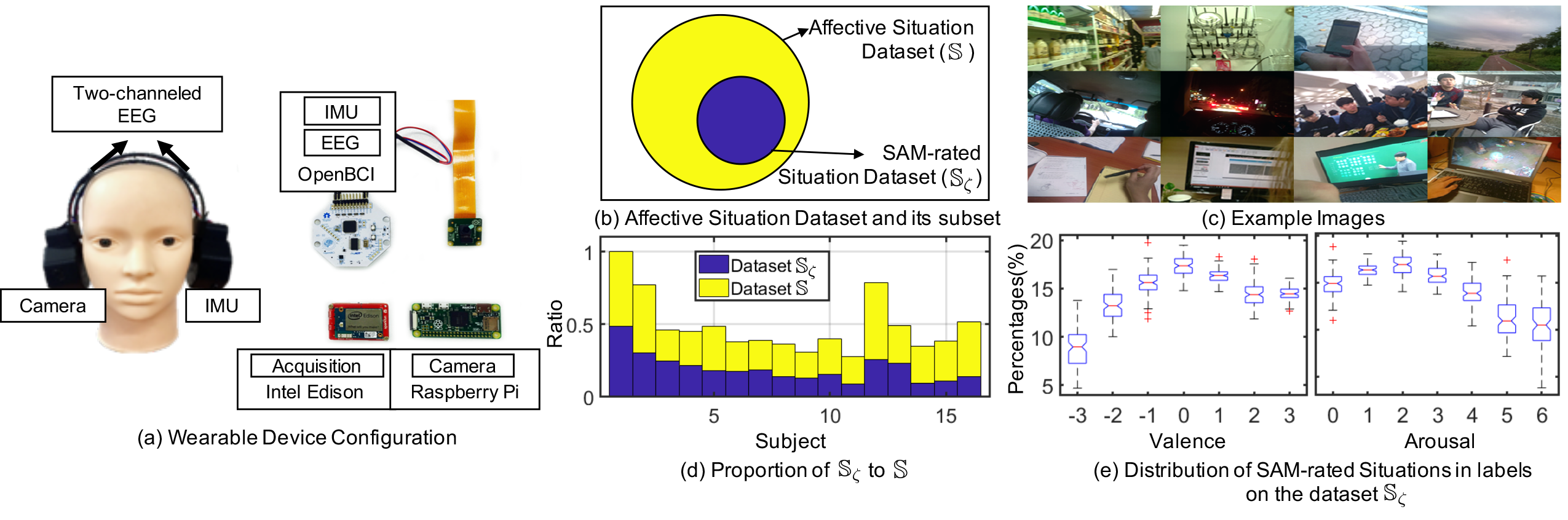}
    \caption{Wearable device configuration and overview of the Affective Situation Dataset. (a) Wearable device configuration. The location of two electrodes (F3, F4) on the 10-20 international system. (b) Affective Situation Dataset $\mathbb{S}$ and its subset $\mathbb{S}_{\zeta}$ which contains SAM-rated situations. (c) Example images from the dataset $\mathbb{S}$. (d) Proportion of the subset $\mathbb{S}_{\zeta}$ and the dataset $\mathbb{S}$. $N_s = 378$ for the subject 1. (e) Distribution of SAM-rated situations in valence and arousal labels on the subset $\mathbb{S}_{\zeta}$.}
    \label{fig:AffectiveSituationOverview}
\end{figure*}
\subsection{Device Configuration}
We designed a simple, easily wearable device (\figurename~\ref{fig:AffectiveSituationOverview}a) such that users could act freely in everyday situations while the device simultaneously, correctly collects their emotions. Since human affect is sophisticated and subtle, it is vulnerable to personal, social, and contextual attributes. The noticeability and visibility of wearable devices could elicit unnecessary and irrelevant emotions, while recording of human affect should be unobtrusive when measured in the natural environment. To design an unobtrusive device, we imitated the design of existing easy-to-use wireless headsets. We note that the term ``unobtrusive device'' means that it is not easily noticed or does not draw attention to itself; it does not imply that our device aims to be small or concealable. This easy-to-use device provides comfort and performance to users during long-term activities.

Our device consists of multimodal sensors to capture various affects surrounding daily life as follows:
\begin{itemize}
\item Frontal Camera for Collecting Visual Content: Visual information has been widely used to detect situations faced by an experimental participant. Analysis of scenes and activities in camera images has provided understanding of this contextual information. Hence, in our system, a small frontal viewing camera with a 30 fps sampling rate was used to record the images.
\item Small Physiological Sensor to Capture Human Affect: Patterns of physiological changes have been increasingly analyzed in the context of affect recognition. To evaluate the reliability of the affective labels predicted by our system, we analyzed the distinctiveness of physiological signals as categorized by the predicted labels. We used a two-channel EEG sensor for the left and right hemispheres, with sampling rates of 250 Hz using OpenBCI, a tool that has been applied successfully in several works~\cite{kaongoen2017two, kaongoen2017novel}.
\end{itemize}

\subsection{Experimental Procedure}
The participants were 13 male and three female students aged 22-35 (27.3$\pm$4.53) years. Participants evaluated our system in a real-world experiment related to school life. They performed more than one common task of a university student, such as taking/teaching classes, conducting research, or having discussions with colleagues. Participants were required to wear our device for 6 hours per day in their daily work environment, for up to 45 days, with \$10 compensation per day.

Participants were asked to engage in free, normal activity over the course of their days. While wearing the device, affective situations are constructed and labeled as pairs of valence and arousal ratings on an affective curve. The modeling of affective situations and their representation as affective curves are respectively described in Sections 3 and 5.1. To evaluate the performance of our system, the participants performed self-assessment of their valence and arousal levels in relation to the affective situations using the web-based SAM, scaled from 0 to 6 for arousal and -3 to 3 for valence. They were asked to rate their feelings spontaneously each day if they had encountered any situation where a certain visual content elicited a specific feeling. In our work, visual content is gathered by the proposed wearable device--book, coffee cup, media device including cell phone, research paper, or monitor. In addition, every five days, we manually retrieved unrated situations containing the visual content which the participant had chosen previously as emotional stimuli and asked the users to rate their feelings in the situation if they could recall them. This procedure was approved by the authors’ Institutional Review Board (IRB) in Human Subjects Research.

Different modalities in our wearable device should theoretically start recording at the same time. In practice, however, sensor measurements are sometime missed for reasons such as battery status or user's mistake. For these reasons, in our implementation we interpolate the missed sensor measurements, so that all measurements are available at the time an image capture begins.

\subsection{Affective Situation Dataset}
\begin{table}[t!]
    \renewcommand{\arraystretch}{1.3}
    \caption{Overview of the dataset contents}
    \label{table:OverviewDataset}
    \centering
\begin{tabular}{ll}
\hline 
\hline 
Number of participants & 16 \\
\hline 
Avg. number of sitations ($\mathbb{S}$)   & 184.5 \\
\hline
Avg. number of SAM-rated situations ($\mathbb{S}_{\zeta}$)   & 74.25 \\
\hline
Avg. number of durations (minute) per situation  & 17.4 \\ 
\hline
Rating values & Valence: -3 to 3 \\ & Arousal: 0 to 6 \\
\hline
Recorded signals & 2-channel EEG \\ & Frontal images \\ & Accelerometer \\
\hline
\end{tabular}
\end{table}
The Affective Situation Dataset $\mathbb{S} = (\mathcal{S}^1, \ldots , \mathcal{S}^{N_s})$ is a set of affective situations collected using the above procedure. Subset $\mathbb{S}_{\zeta} \subset \mathbb{S}$, which has a pair of valence ($\mathcal{V}$) and arousal ($\mathcal{A}$) ratings rated by the SAM, consists of the SAM-rated Situation Dataset. The affective labels ($\mathcal{V},\mathcal{A}$) of situations were used as ground truth to estimate the parameters of our system and evaluate its performance. The duration $T$ of all situations was determined manually by three annotators, who spent 2.4 ($\pm$1.2) minute per situation. The inter-rater reliability was measured using interclass correlation (ICC); the result was 0.78. The average duration from the three annotators was ultimately used for each situation. \figurename~\ref{fig:AffectiveSituationOverview} summarizes the affective situation dataset. \figurename~\ref{fig:AffectiveSituationOverview}b shows some example images for the dataset $\mathbb{S}$ from our real-world experiments. The distribution of the SAM-rated situations, which consist of $\mathbb{S}_{\zeta}$ in different ratings for valence and arousal, is shown in \figurename~\ref{fig:AffectiveSituationOverview}e. Table~\ref{table:OverviewDataset} gives an overview of the dataset contents.

\section{Experiments Using the Affective Situation Dataset}
With the Affective Situation Dataset, we evaluated the performance of our system for labeling emotions compared with the labels rated by the SAM. Furthermore, the distinctiveness of EEG signals categorized by different labels was also evaluated by comparing it with other state-of-the-art methods.

\subsection{Parameter Setting of A-Situ}
To model affective situations and represent them as affective curves, the parameters introduced in Section 3 for each participant were set through a five-fold cross-validation scheme, except $\lambda_2$ and $\lambda_3$. These parameters were set to 2 and -1, indicating that the contentment component has a minimum value of $-1/l_{max}$ at the first frame in a situation. $m_{max}$, $l_{max}$, $r_{max}$, and $o_{max}$ are determined by the maximum values during every five days. 
\begin{figure}[t]
\centering
\subfloat[$\lambda_1$ for valence]{\includegraphics[width=0.5\columnwidth]{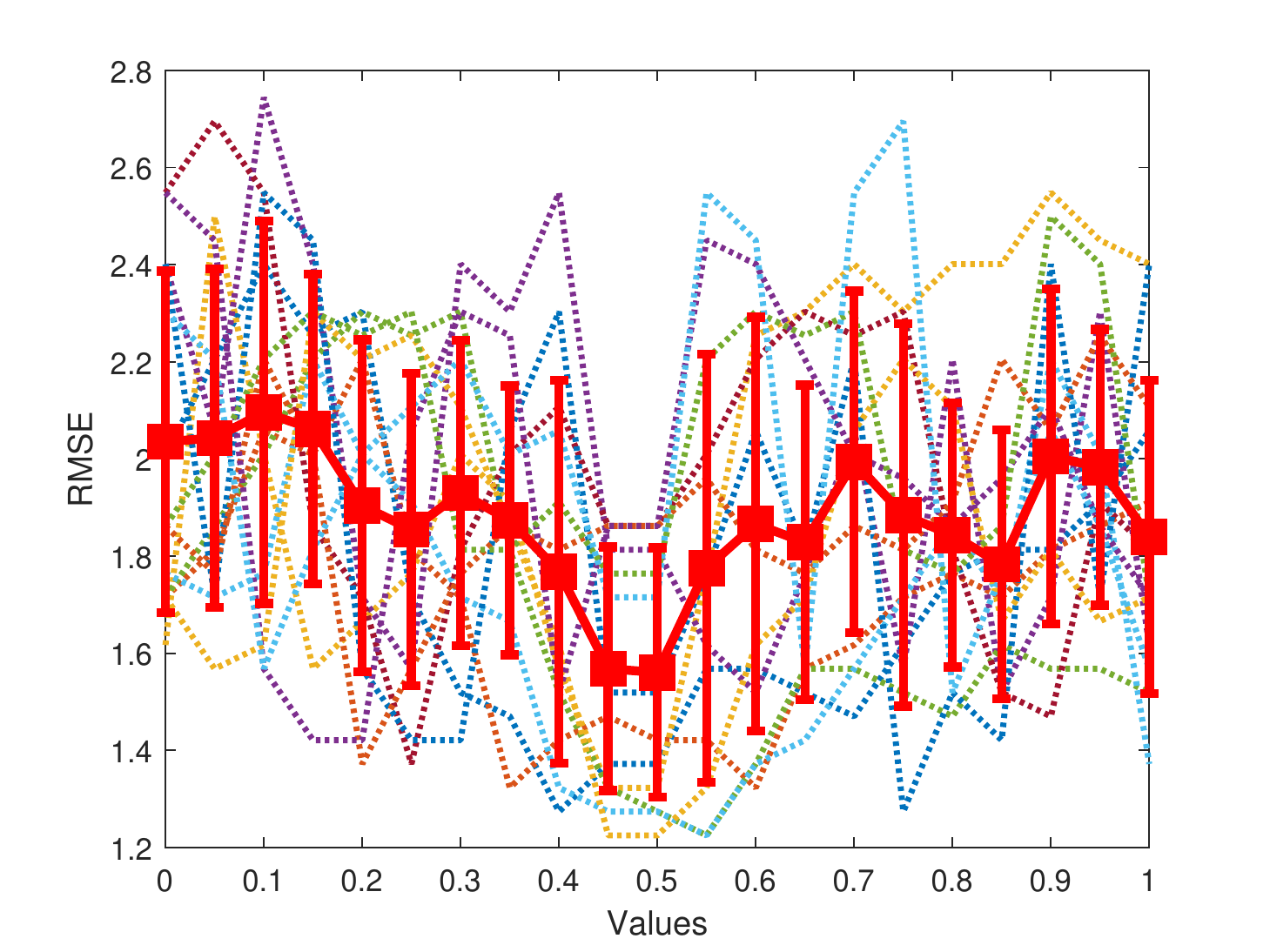}
\label{fig:para_lambda3_valence}}
\subfloat[$\lambda_1$ for arousal]{\includegraphics[width=0.5\columnwidth]{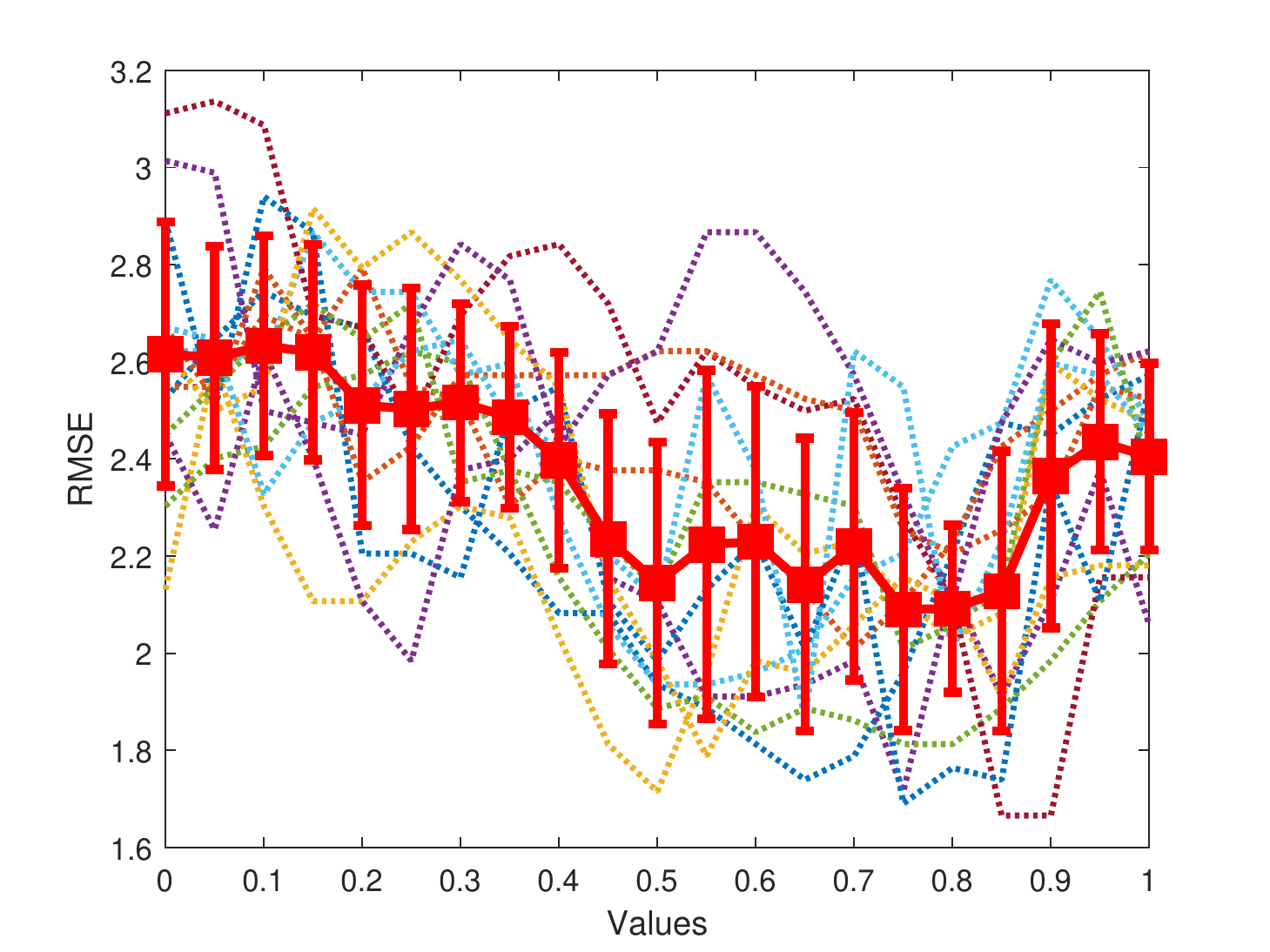}
\label{fig:para_lambda3_arousal}}
\hfil
\subfloat[$\alpha_1$]{\includegraphics[width=0.5\columnwidth]{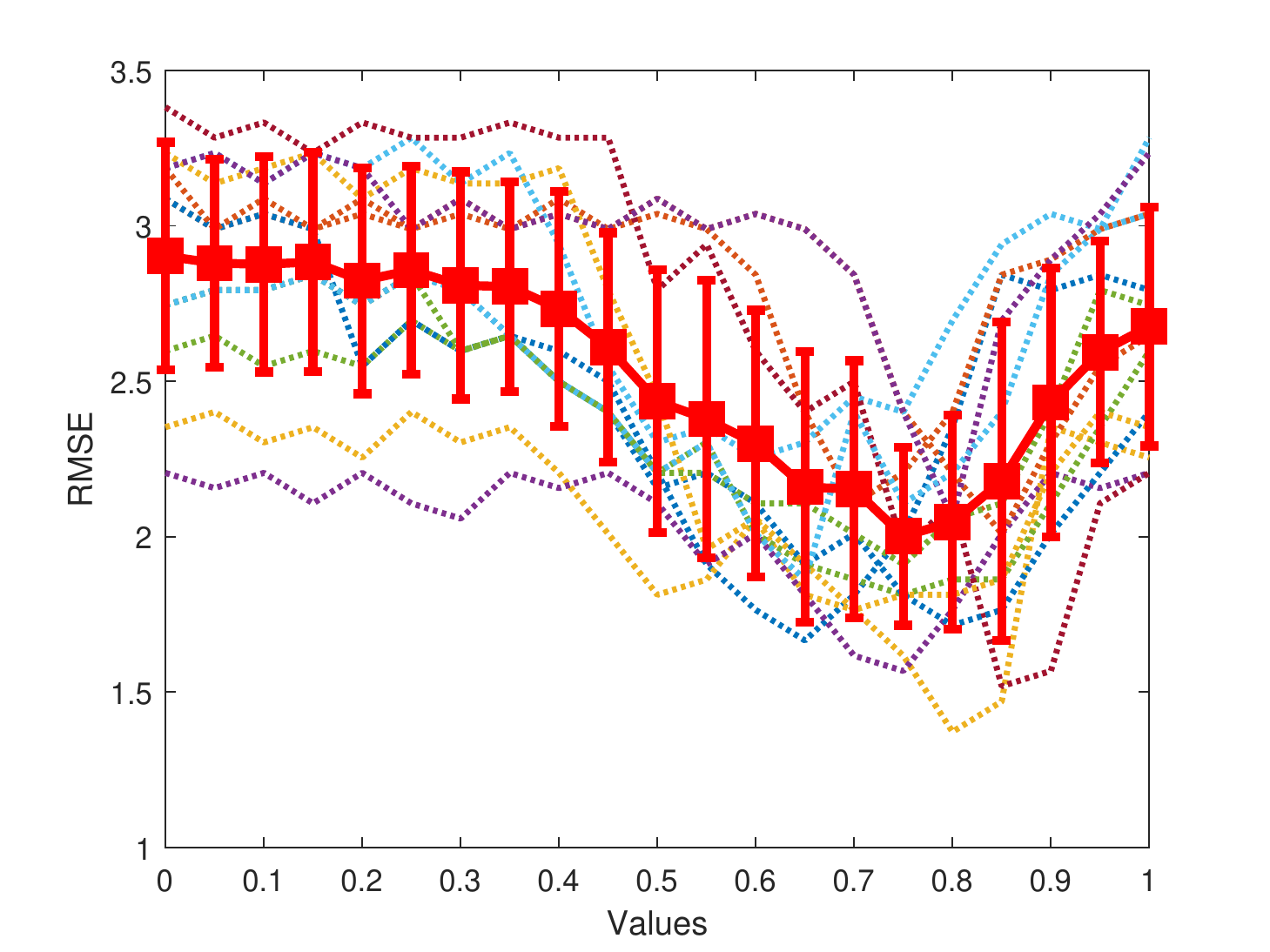}
\label{fig:para_alpha}}
\subfloat[$\nu_1$]{\includegraphics[width=0.5\columnwidth]{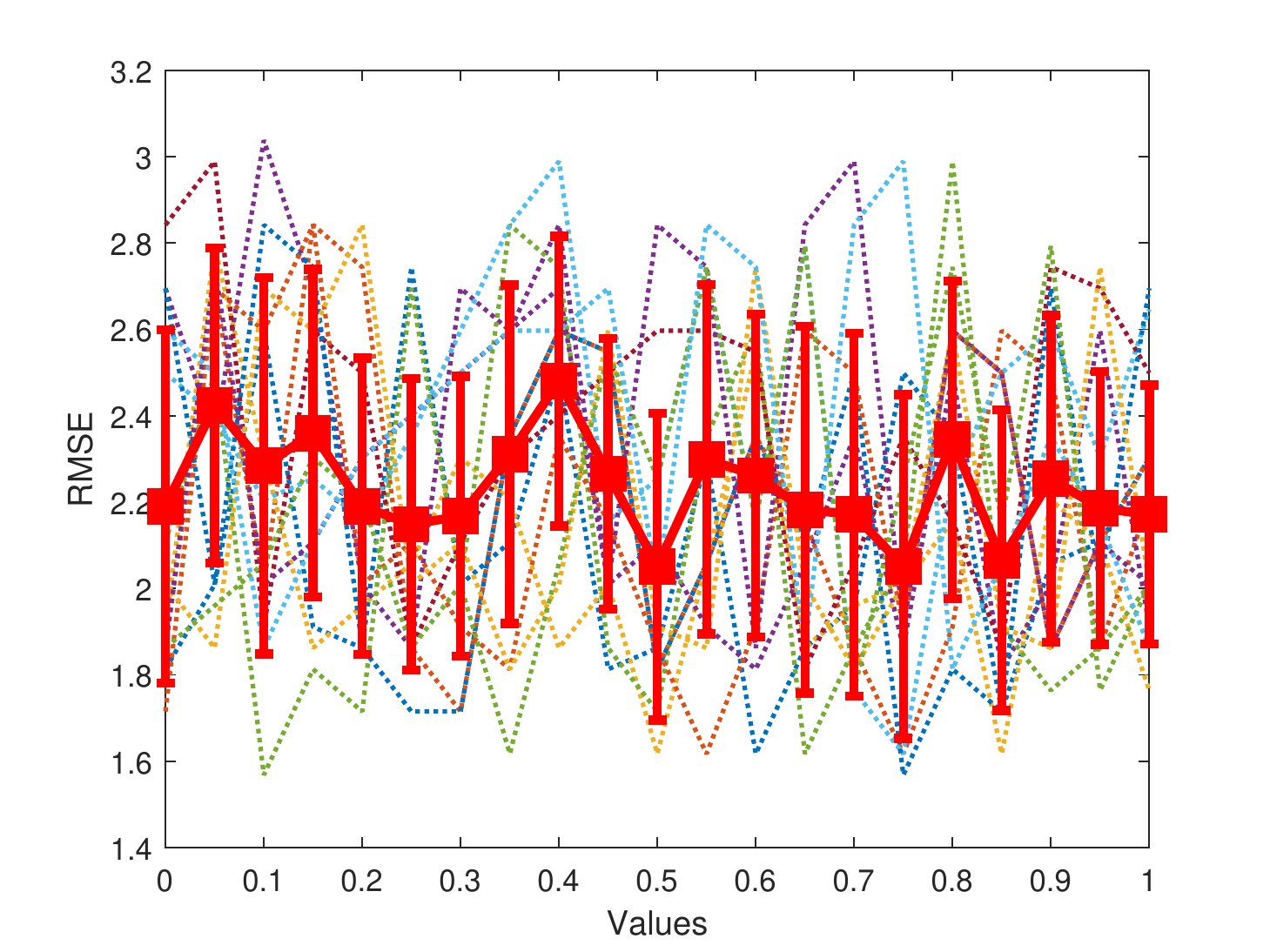}
\label{fig:para_nu}}
\caption{Accuracy results of valence and arousal ratings by the proposed system with respect to (a) and (b)$\lambda_1$, (c)$\alpha_1$, and (d)$\nu_1$}
\label{fig:ParaSetting}
\end{figure}

\figurename~\ref{fig:ParaSetting} shows the results from the five-fold cross-validation in terms of valence and arousal rating distances between the SAM and A-Situ with respect to different parameters $\alpha_1$, $\lambda_1$, and $\nu_1$. Because an affective curve represents affective dynamics in a spatiotemporal situation and consists of multiple pairs of affective labels, it is difficult to directly compare them with a pair of SAM ratings, in which the valence and the arousal ratings are a pair of discrete values for representing an emotion in the same situation. For comparison, we calculate root-mean-squared-errors (RMSEs) of a pair of affective labels scaled 0 to 6 over all participants, as follows:

\begin{small}
\begin{equation}
\label{eq:RMSE}
RMSE = \sqrt{\frac{1}{N_s}\sum_{i=1}^{N_s}(\hat{y}_i - y_i)^2},
\end{equation}
\end{small}where $N_s$ is the numbers of situations in $\mathbb{S}_{\zeta}$, $\hat{y}_i$ is the mean value of pairs of the predicted affective labels ($\mathcal{V}$, $\mathcal{A}$), and $y_i$ is a pair of ground truth labels in situation $i \in \mathbb{S}_{\zeta}$. Note that 0 represents negative, 3 neutral, and 6 positive valence ratings; and 0 represents neutral, 3 represents low, and 6 represents high arousal ratings. 

While $\alpha_w$ and $\lambda_w$ ($w=1,2$) determine the arousal value $\mathcal{A}(\mathcal{S}^i_t)$, the valence value $\mathcal{V}(\mathcal{S}^i_t)$ is determined by the parameters $\lambda_w$ and $\nu_w$. The parameter $\lambda_1$ determines when the contentment component $l(\mathcal{S}^i_t)$ in (\ref{eq:adaptation}) becomes zero; this directly affects the sign of the valence value $\mathcal{V}(\mathcal{S}^i_t)$ and increment of the arousal value $\mathcal{A}(\mathcal{S}^i_t)$. With a smaller $\lambda_1$ value, the contentment component $l(\mathcal{S}^i_t)$ becomes zero and the valence sign $\mathcal{V}(\mathcal{S}^i_t)$ becomes positive at an earlier $t$. As shown in \figurename~\ref{fig:para_lambda3_valence}, the highest performance on valence ratings was 0.5, and two other points, 0.25 and 0.75, were the next highest parameters for valence ratings. Conversely, the distances between arousal ratings are minimized after 0.5 for most participants, as shown in \figurename~\ref{fig:para_lambda3_arousal}. The parameter $\alpha_1$ and its counterpart $\alpha_2(=1-\alpha_1)$ determine the level of arousal in terms of the motion and contentment components. The results reveal that the proposed system has the best performance for arousal ratings at between 0.7 and 0.8 for most participants (see \figurename~\ref{fig:para_alpha}) with respect to $\alpha_1$. Distances beyond this range remain almost the same when $\alpha_1$ is greater than 0.85 and less than 0.4. The parameter $\nu_1$ and its counterpart $\nu_2(=1-\nu_1)$ determine the level of valence. With a large value of $\nu_1$, the resulting valence may not properly represent the emotion associated with approach-withdrawal behaviors. As shown in \figurename~\ref{fig:para_nu}, the performance of our proposed system fluctuates significantly when the parameter $\nu_1$ is varied. We set $\alpha_1$ and $\nu_1$ to be 0.75 and 0.5 for all participants and $\lambda_1$ was chosen from 0.25, 0.5, and 0.75 to yield the minimum distance for each individual in each of the following experiments. 

\subsection{EEG Preprocessing and Setup for Classification}
As a preprocessing step, high-pass filtered with a 2-Hz cutoff frequency using the EEGlab toolbox and the same blind source separation technique for removing eye artifacts were applied. A constrained independent component analysis (cICA) algorithm was applied to refine the signal removing motion artifacts~\cite{breuer2014constrained}. The cICA algorithm is an extension of ICA and has been applicable in cases in which prior knowledge of the underlying sources is available~\cite{lu2005approach}. 

EEG signals are vulnerable to motion artifacts~\cite{uriguen2015eeg}. Rather than separating and removing motion artifacts in EEG signals occurred by body movement~\cite{li2017discriminative, daly2013automated}, we developed a strategy to get better-quality EEG signals by abandoning signals highly correlated with motion artifacts. To execute this strategy, we subdivided EEG signals into two groups separated by the accelerometer data $\mathcal{E}^i_t$ ranged from 0 to 1 in (\ref{eq:motion2}). From each of the two groups, we extract the following EEG features: 1) mean power, 2) maximum amplitude, 3) standard deviation of the amplitude, 4) kurtosis of the amplitude, and 5) skewness of the amplitude. These features are metrics to describe the key characteristics of clean EEG~\cite{daly2012does}. After representing the features in two-dimensional space using principal component analysis (PCA), we compute the Bhattacharyya distance between the two groups over the two-dimensional space. The optical $G(\cdot)$ is determined as a differentiator between the clean EEG and the contaminated EEG, based on the maximum distance between the two groups. 

Recent studies on extracting EEG-based features in emotion recognition have categorized these features into three domains: time, time–frequency, and frequency~\cite{alarcao2017emotions}. Among these, frequency domain features have been the most popular, assuming that the signal remains stationary for the duration of a trial. Hence, we used frequency domain features introduced in \cite{jenke2014feature}:higher-order spectra (HOS) and power spectral density (PSD) features in different frequency bands. HOS features have been used to analyze human emotion as a spectral representation of higher-order moments or cumulants of a signal~\cite{jenke2014feature}. Specifically, we used the mean of bicoherence in four frequency bands—theta (4–7 Hz), alpha (8–13 Hz), beta (14–29Hz), and gamma (30–45 Hz)—to study the efficacy of affective labels to categorize EEG signals. Bicoherence is the normalized bispectrum of a signal $x(t)$. Signals are divided into 1-s non-overlapping segments. Within each segment, data are Hanning windowed and Fourier transformed. Then, the bispectrum $B(\omega_1, \omega_2)$ is mathematically defined as
\begin{small}
\begin{equation}
\label{eq:bic1}
B(\omega_1, \omega_2) = X(\omega_1)X(\omega_2)X^*(\omega_1 + \omega_2),
\end{equation}
\end{small}where $X(\omega)$ is the Fourier transform of the signal $x(t)$ and $X^*(\omega)$ is its complex conjugate. Note that the bispectrum preserves phase information of the different components of the signal $x(t)$. Two frequency components $X(\omega_1)$ and $X(\omega_2)$ are phase coupled when there exists a third component at a frequency of $\omega_1 + \omega_2$. The bicoherence $b_c(\omega_1, \omega_2)$ is defined as 
\begin{small}
\begin{equation}
\label{eq:bic2}
b_c(\omega_1, \omega_2) = \frac{|B(\omega_1, \omega_2)|}{\sqrt{P(\omega_1)P(\omega_2)P(\omega_1+\omega_2)}},
\end{equation}
\end{small}where $P(\omega_i)$ is the power spectrum at $\omega_i$. It quantifies the extent of phase coupling between two frequency components. The resulting frequency resolution is 1 Hz on at both the $\omega_1$ and $\omega_2$ axis. The mean magnitude of $b_c(\omega_1, \omega_2)$ in the four frequency bands is computed as
\begin{small}
\begin{equation}
\label{eq:bic3}
b_c^{avg}(q_1,q_2) = \frac{1}{L}\sum_{q_1}\sum_{q_2}b_c(q_1, q_2),
\end{equation}
\end{small}where $q_1$ and $q_2$ are frequency bands and $L_{(q_1,q_2)}$ is the number of frequency components in $q_1$ and $q_2$. Power features of the PSD are estimated using Welch’s method~\cite{jenke2014feature} and divided into the four frequency bands. The $b_c^{avg}$ and the mean power of the four frequency bands are used to analyze the correlates of the affective labels with EEG signals.

\subsection{Evaluated Methods}
\begin{figure}[t]
    \centering
    \includegraphics[width=0.8\columnwidth]{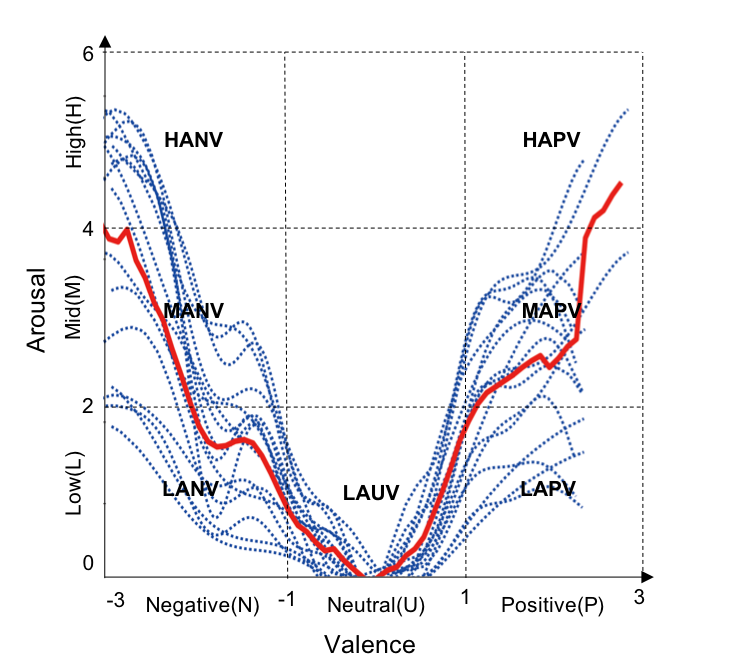}
    \caption{Affective states subdivided by low(LA), mid(MA), and high(HA) arousal and negative(NV), neutral(UV), and positive(PV) valence ratings, for all participants over different affective states in the dataset $\mathbb{S}$. Dashed lines indicate individual participants, and a solid red line is the mean curve of all participants.}
    \label{fig:AffectiveStates}
\end{figure}
The efficiency of affective labels provided by our system to discriminate different states in EEG-based emotion recognition was evaluated by subject-dependent classification performance using HOS and PSD features through a five-fold cross-validation scheme for all participants. As shown in \figurename~\ref{fig:AffectiveStates}, we subdivided affective labels over the valence–arousal space into low, mid, and high states for arousal and negative, neutral, and high state for valence. We should note that the results of the ANOVA tests for the bicoherence magnitudes and the PSD in the four frequency bands of the affective states were low $p$-values (lower than 0.05), except the beta frequency band ($p$=0.0679). The $p$-values resulted from the bicoherence magnitudes in all frequency bands, and PSD in the theta, alpha and gamma frequencies indicated that the three frequency bands appear to be significantly different from emotional states. These results imply that PSD and bicoherence can be used effectively as physiological features to classify emotions.

For the classification process, we choose two classifiers: a support vector machine (SVM) and a ConvLSTM, both of which have been used widely in emotion recognition~\cite{alarcao2017emotions}. For SVMs, we extract the PSD and bicoherence features $b_c(\omega_1, \omega_2)$ in the four frequency bands, use mutual information for feature selection, and take the selected features as input for classification. For ConvLSTMs, the PSD features in the four frequency bands are fed into ConvLSTMs, as in~\cite{kim2018deep} to classify the affective states. To compare classification results, the following models are trained by the two classifiers and evaluated on Affective Situation Dataset $\mathbb{S}$.

\subsubsection{Baseline I}
The model is trained on the dataset $\mathbb{S}_{\zeta}$; which labels in affective situations were rated by the SAM. To compare the performance with the other two methods, the model is evaluated on the datasets of both $\mathbb{S}_{\zeta}$ and $\mathbb{S}$.
\subsubsection{Baseline II}
For the algorithm, like \cite{hanjalic2005affective}, we replace shot lengths with $T_i$ for situation $i$. The sound energy and pitch-average components are excluded from computation of affective labels, since dataset $\mathbb{S}$ does not include any sound. The model is trained and evaluated on both $\mathbb{S}_{\zeta}$ and $\mathbb{S}$. Since the model only rates arousal labels, evaluation is carried out to classify affective states associated with arousal: low-arousal (LA), mid-arousal (MA), and high-arousal (HA) states.
\subsubsection{Our Proposed Model}
Our proposed model is trained, and evaluated on both $\mathbb{S}_{\zeta}$ and $\mathbb{S}$, for which labels were computed by A-Situ in Section 3. 

\subsection{Experimental Results}
\begin{figure}[t]
    \centering
    \includegraphics[width=0.8\columnwidth]{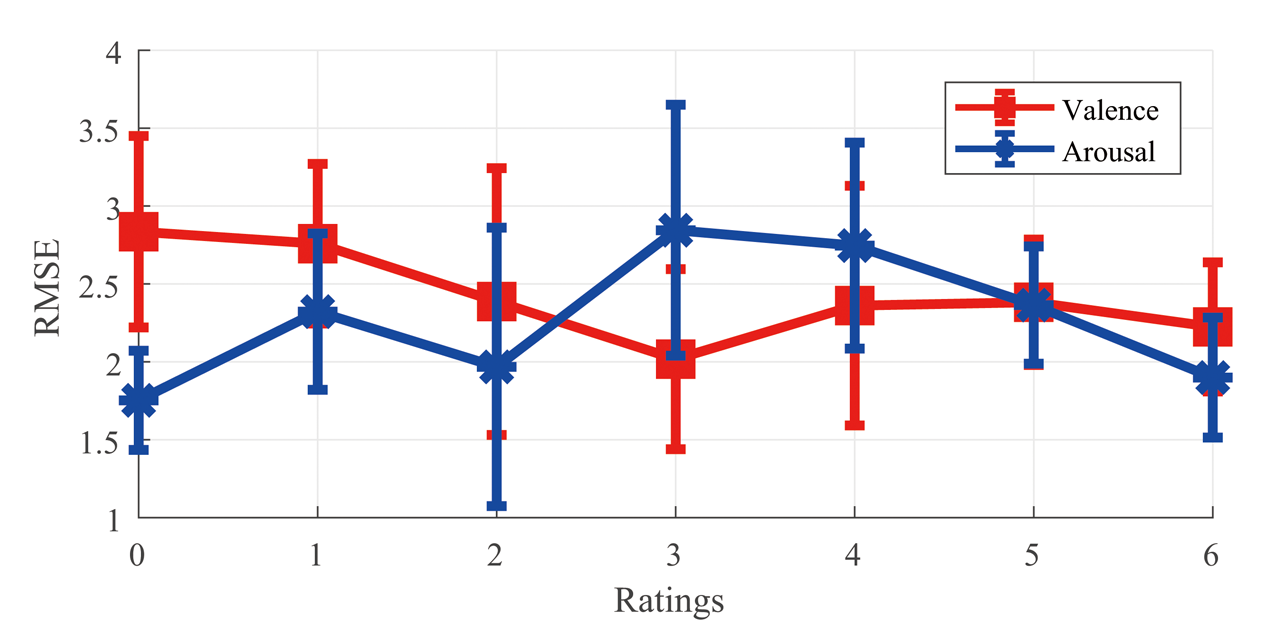}
    \caption{Mean RMSEs of all SAM-rated situations on dataset $\mathbb{S}_{\zeta}$ between predicted labels and ground truth labels, scaled 0 to 6 over valence (red) and arousal (blue) dimensions. Note that 0 represents negative, 3 neutral, and 6 positive valence ratings, while 0 represents neutral, 3 low, and 6 high arousal ratings.}
    \label{fig:MeanRMSEsofallSAMratedSituations}
\end{figure}

\figurename~\ref{fig:MeanRMSEsofallSAMratedSituations} shows the mean RMSEs of all SAM-rated situations on dataset $\mathbb{S}_{\zeta}$ between labels predicted by the proposed system and ground truth labels as rated by the SAM. Note that 0 represents negative, 3 neutral, and 6 positive valence ratings, while 0 represents neutral, 3 low, and 6 high arousal ratings. The mean accuracies for valence and arousal ratings between the two labels were respectively 2.42 ($\pm$0.59) and 2.27 ($\pm$0.7), equivalent to 65.42\% and 67.57\% in terms of normalized RMSE. In both cases, neutral ratings had smallest errors; while higher ratings on arousal had more similarity, negative ratings on valence had larger errors with larger standard deviations. We should affirm that A-Situ does not aim to evaluate affective labels at the same precision level as the SAM ratings; instead, the primary purpose of the proposed system is to provide reliable emotion labels associated with physiological characteristics derived from psychological behaviors.

\begin{figure}[t]
    \centering
    \includegraphics[width=0.8\columnwidth]{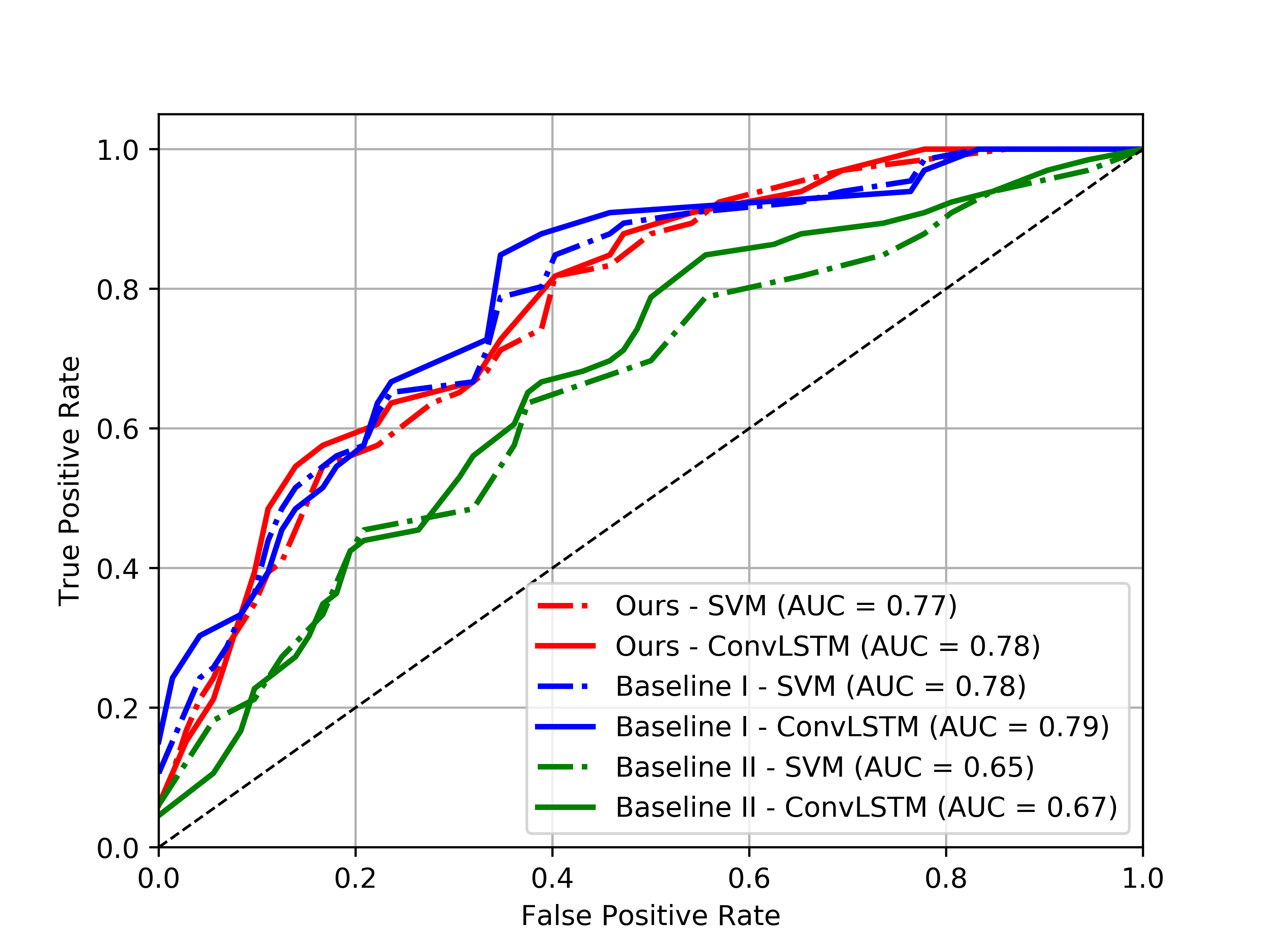}
    \caption{Comparisons of classification results between the proposed system and Baselines I \& II methods on the dataset $\mathbb{S}_{\zeta}$.}
    \label{fig:classificationComparison1}
\end{figure}

\begin{figure}[t]
    \centering
    \includegraphics[width=0.8\columnwidth]{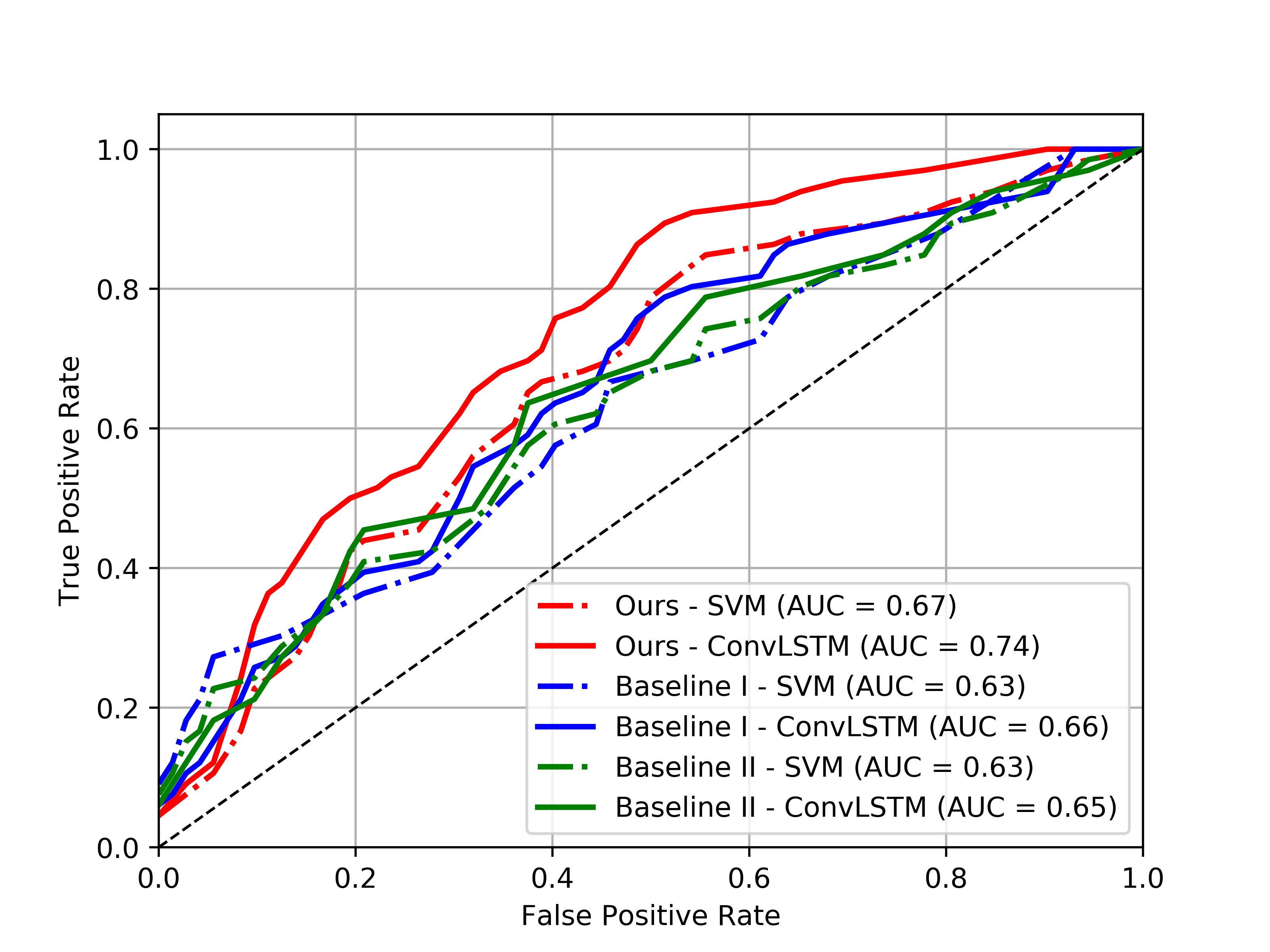}
    \caption{Comparisons of classification results between the proposed system and Baselines I \& II methods on the dataset $\mathbb{S}$.}
    \label{fig:classificationComparison2}
\end{figure}

\begin{figure}[t]
    \centering
    \includegraphics[width=0.8\columnwidth]{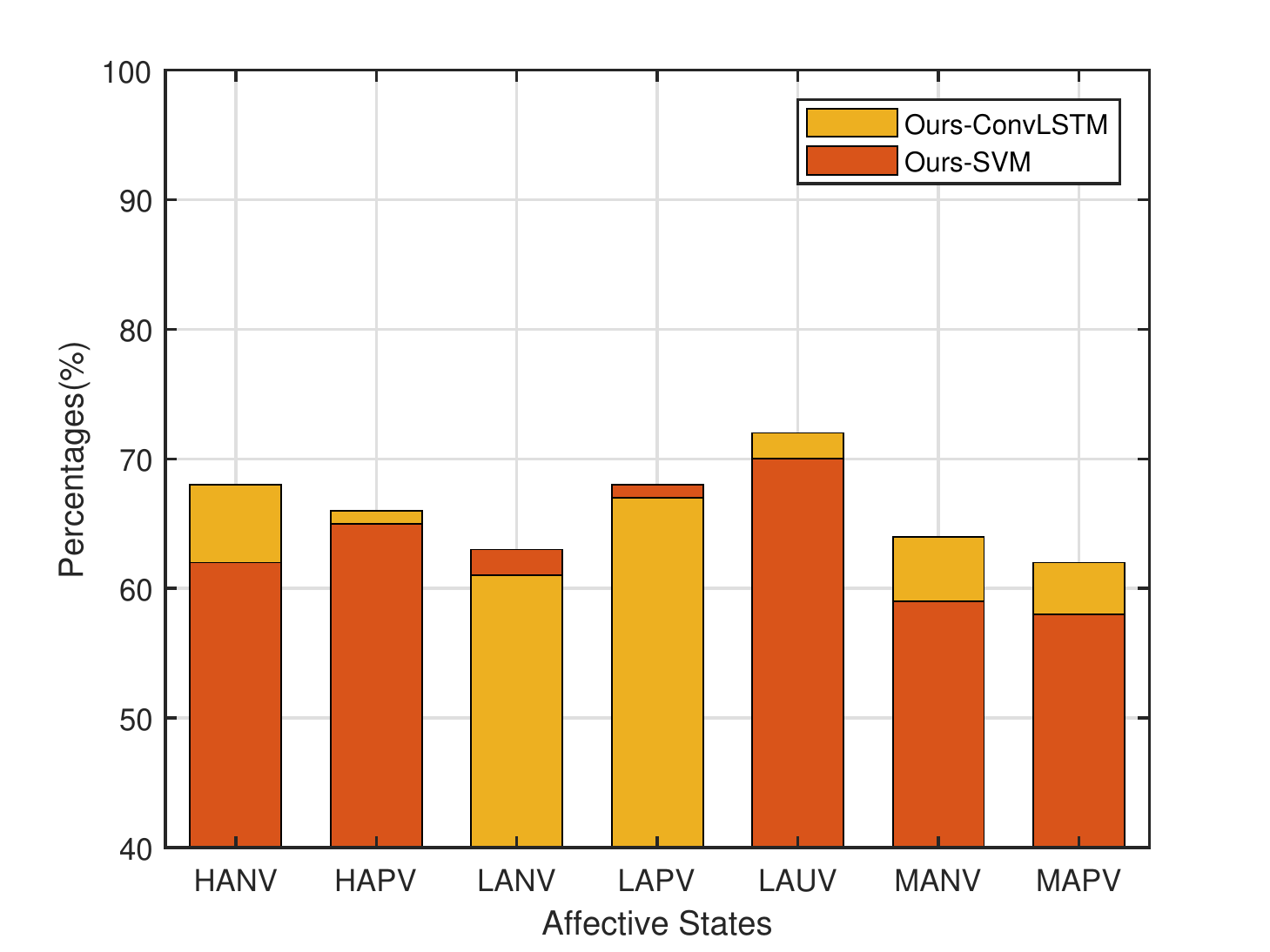}
    \caption{Mean classification results of seven affective states over V-A emotion space for all participants on the dataset $\mathbb{S}$.}
    \label{fig:statesClassificationResults}
\end{figure}

\figurename~\ref{fig:classificationComparison1} and \ref{fig:classificationComparison2} show the evaluation results for the two sets of affective situations. For dataset $\mathbb{S}_{\zeta}$, as shown in \figurename~\ref{fig:classificationComparison1}, our system performed comparably to Baseline I. Although it achieved slightly worse results than Baseline I when 0.2$<$FP$<$0.6, these two methods perform equally well overall on the dataset. These results can be attributed to the fact that the labels provided by our proposed system categorize EEG signals associated with different emotions. Although the predicted labels obtained from our system have different interpretation from the SAM ratings by the Baseline I for rating real-world situations (see \figurename~\ref{fig:MeanRMSEsofallSAMratedSituations}), the classifiers based on our system achieve similar performance to those based on the SAM ratings. In contrast, the results of the Baseline II method are the worst for all cases; this can be explained by noting that the use of optical flow-based motion components alone has less discriminative power to classify physiological patterns in various situations.

\figurename~\ref{fig:classificationComparison2} shows the ROC curves for dataset $\mathbb{S}$. Overall, the proposed system performed favorably in classifying emotional states with higher area under the curve (AUC) than any of the baseline methods. Although ConvLSTMs increased the distinctiveness in order to classify EEG signals labeled by Baseline I, the two methods are less discriminative than the proposed method in terms of AUC. This superior performance by our system demonstrates the effectiveness of the proposed system for overcoming intra-subject variability in EEG signals. The classifiers more reliably learn physiological patterns in EEG signals associated with affective states predicted by our model than do those rated by the SAM. Furthermore, these results imply that the proposed system performs robustly in real-world environments with their many different possible situations.

\figurename~\ref{fig:statesClassificationResults} shows accuracy across the classification of the seven affective states (HANV, HAPV, LANV, LAPV, LAUV, MANV, and MAPV, as shown in \figurename~\ref{fig:AffectiveStates}) subdivided by our system's predicted labels for dataset $\mathbb{S}$. The results from all participants averaged 63.57\% and 65.71\% for SVMs and ConvLSTMs, respectively. The LAUV state archived the highest accuracy when using both classifiers, implying that when participants are in the LAUV, experiencing calm, relaxed feelings, they have distinct activation patterns from when they are in the other states. In terms of performance by the two classifiers, ConvLSTMs achieved stable results with higher accuracies and lower standard deviations over all states. For the MANV, MAPV, and HANV states, our SVMs yielded lower accuracies than did the results from ConvLSTMs. This can be attributed to the fact that the percentages of affective situations rated as low arousal (between 0 and 3) were higher than the others in dataset $\mathbb{S}$ (see \figurename~\ref{fig:AffectiveSituationOverview}), which could lead to SVMs becoming overfit and overconfident to some labels, such as low arousal states.

\section{Case Study on Affective Situation Dataset}

\begin{figure}[t]
    \centering
    \includegraphics[width=1\columnwidth]{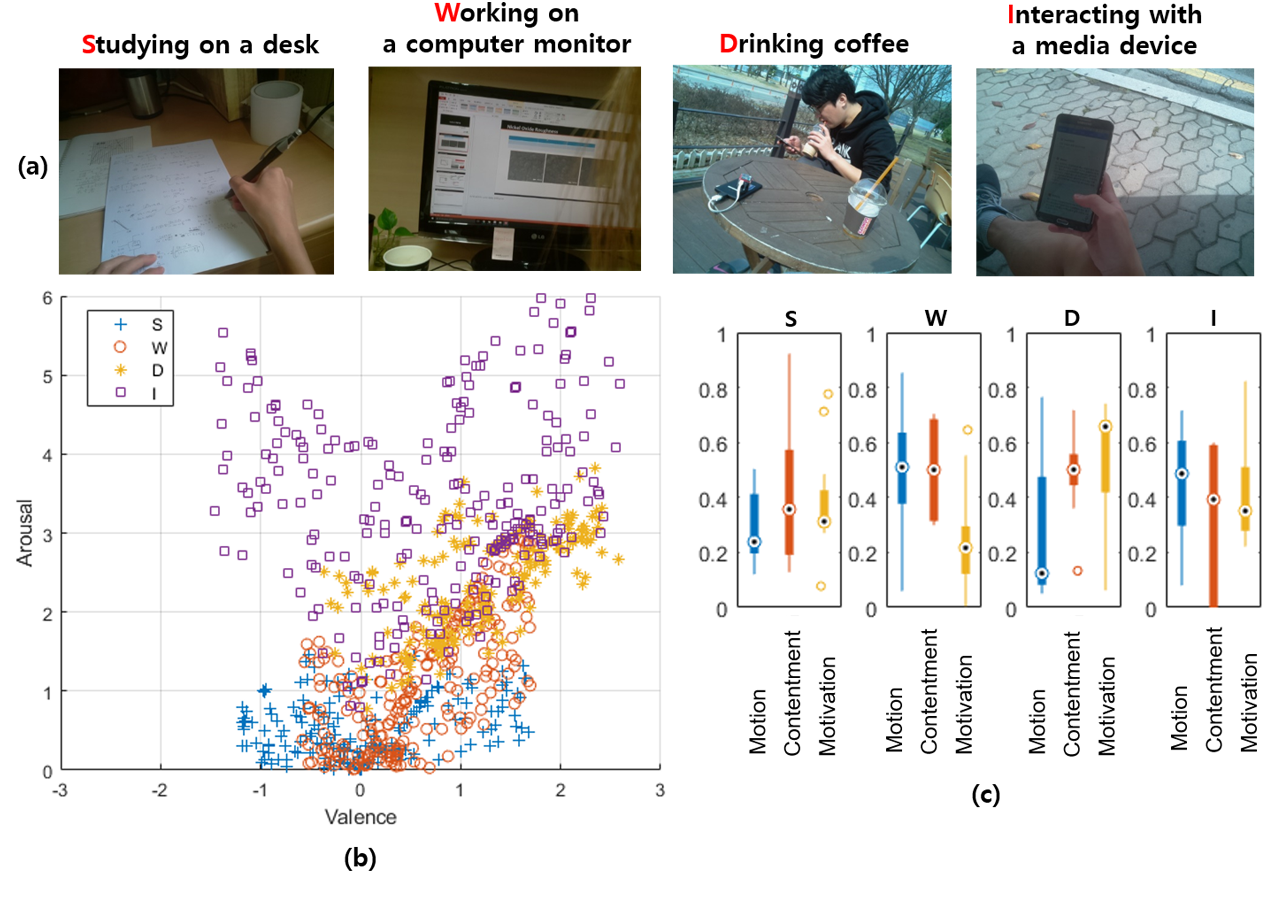}
    \caption{The four most frequent affective situations. Example images, accumulated emotion points over valence and arousal points, and the means of the motivation, motion, and contentment components for all participants.}
    \label{fig:freqSitu}
\end{figure}
The proposed A-Situ provides affective labels underlying physiological characteristics associated with psychological phenomena. Since the framework outputs a set of affective labels in a spatiotemporal situation, pairs of labels on the affective curve contain emotional traces in response to the affective content of a situation. \figurename~\ref{fig:AffectiveStates} shows affective curves created by combining the arousal and valence curves in (\ref{eq:VAmodel1}) and (\ref{eq:VAmodel3}). Each curve represents the emotional representation of affective situations in the everyday life of a participant. The parabolic shape of the mean curve covers the V–A emotion space, except for some emotions characterized by neutral valence and high-level arousal.

To demonstrate the effectiveness of the model, we show some interesting cases that involve analyzing physiological characteristics. We choose the four most frequent situations: ``Working on a computer'', ``Studying at a desk'', ``Drinking coffee'', and ``Interacting with a media device'' on the dataset $\mathbb{S}$. \figurename~\ref{fig:freqSitu} shows example images, accumulated valence and arousal labels over valence–arousal dimension, and the three components.

The situation ``Studying at a desk'' drew affective curves around the valence and arousal values between -1 and 1 and between 0 and 1.3, respectively; these low scores were due to results from the motion and motivation components rather than the contentment component (see \figurename~\ref{fig:freqSitu}c). This phenomenon indicates that most participants in this situation spent longer sitting stationary than in other situations to keep concentrating while studying. This activity in the situation yielded lower motivation and motion components but a higher contentment component. Negative affect occurs when participants have low motivation and contentment components, leading them to stop studying and leave the situation earlier than usual. Such thwarted goals incur negative feeling such as frustration.

The situation ``Working on a computer monitor'' drew similar affective curves to ``Studying at a desk,'' but had larger values for the motion component than the latter. This indicates that interaction with a computer monitor, such as exploring/searching websites, lead to larger motion changes in display than ``Studying at a desk'', but smaller than the other activities, with correspondingly higher/lower arousal values.

The situation ``Drinking coffee'' includes activities whose affective curves were affected by the motivation and contentment components. When participants stayed in their circumstances and drank coffee while interacting with other factors, the two components had high values, resulting in a high valence score. For example, approaching (drinking) a cup of coffee, reading a book, and hanging out with friends led to increased values of the two components, which resulted from the movements of either hand while approaching the coffee cup or other movements during the long sequence of the situation. Since this personalization determines the degree of the valence score, this score was highly variable, with a standard deviation of $\pm$0.7.

The situation ``Interacting with a media device'' includes activities where participants interact with several digital media, such as playing PC games, watching YouTube videos, or posting on the social media. This situation had the highest variance in valence and arousal scores, and the contentment and motion components were spread wider than the motivation component. The level of acceptance of frequent motion of objects in media and playing or using them for a long time led to changes in valence and arousal scores. Negative affect such as frustration can occur when participants have low motivation and contentment components, implying loss of interest and leaving the situation earlier than usual when they thwart their own goals, such as through an unexpected loss in a game.

\subsection{Analysis of Physiological Characteristics}
We have shown the efficacy of affective curves as reliable emotion labels. However, this finding is limited to the clarification of affective labels reflecting physiological characteristics; that is, it only shows the distinctiveness of EEG signals associated with the labels, which were produced based on psychological measurements. Therefore, we investigated the statistical relationship between EEG spectral power in the four frequency bands from two electrodes (F3, F4) and the affective labels for bridging the gap between psychological measurements and physiological evidence.

\begin{table}[t!]
    \renewcommand{\arraystretch}{1.3}
    \caption{Mean Spearman's correlation coefficients of valence and arousal ratings with EEG spectral power in the four frequency bands (theta, alpha, beta, and gamma) for all participants.}
    \label{table:VAEEGcorrelation}
    \centering
\begin{tabular}{ccccc}
\hline 
\hline 
& \multicolumn{2}{c}{Valence} & \multicolumn{2}{c}{Arousal}   \\ 
\cline{2-5} 
Frequency & F3 & F4 & F3 & F4 \\
\hline 
Theta       & 0.31($\pm$0.04) & 0.22($\pm$0.11) & 0.12($\pm$0.09) & 0.17($\pm$0.08) \\
\hline
Alpha       & $\textbf{0.37}$($\pm$0.11) & $\textbf{0.26}$($\pm$0.12) & $\textbf{0.22}$($\pm$0.15) & $\textbf{0.26}$($\pm$0.05) \\
\hline
Beta        & 0.04($\pm$0.06) & 0.17($\pm$0.05) & 0.02($\pm$0.14)  & 0.11($\pm$0.05)\\
\hline
Gamma	    & 0.12($\pm$0.08) & 0.21($\pm$0.12) & 0.17($\pm$0.08) & 0.23($\pm$0.09) \\ 
\hline
\end{tabular}
\end{table}
Table~\ref{table:VAEEGcorrelation} indicates that the alpha frequency components have higher correlations to both ratings than the other frequencies. Since our dataset contains participants' hand movements, the motor cortex activation related to the movements can be correlated with either valence or arousal ratings, because Mu rhythms (8–11 Hz) activated by movements in the motor cortex area have strong associations with the alpha frequency components.

\begin{table}[t!]
    \renewcommand{\arraystretch}{1.3}
    \caption{Mean Pearson's correlation coefficients of the three components (motion, motivation, and contentment) with EEG spectral power in the four frequency bands (theta, alpha, beta, and gamma) for all participants.}
    \label{table:VAComponentscorrelation}
    \centering
\begin{tabular}{ccccccc}
\hline 
\hline 
& \multicolumn{2}{c}{Motion} & \multicolumn{2}{c}{Motivation} & \multicolumn{2}{c}{Contentment} \\ 
\cline{2-7} 
Frequency & F3 & F4 & F3 & F4 & F3 & F4 \\
\hline 
Theta       & 0.14 & 0.08 & 0.19 & 0.13 & 0.05 & 0.07 \\
\hline
Alpha       & $\textbf{-0.08}$ & $\textbf{0.01}$ & 0.21 & 0.18 & $\textbf{0.16}$ & $\textbf{0.13}$ \\
\hline
Beta        & 0.05 & -0.07 & -0.08  & -0.05 & 0.03 & 0.09\\
\hline
Gamma	    & 0.27 & 0.21 & -0.12 & 0.03 & 0.11 & 0.05 \\ 
\hline
\end{tabular}
\end{table}
However, this does not imply psychological measurements in hand movements are uncorrelated with physiological changes. To determine if the alpha frequency band is contaminated by hand movements or correlated with psychological measurements in action, we computed correlation coefficients between the power of the four frequency bands and the three components. Table~\ref{table:VAComponentscorrelation} shows the correlation between physiological brain activity and the affective states, for each component. As we described in Section 3, the motion and motivation components may potentially be influenced by the movements; however, only the motivation component reflects a subject's approach behaviors in psychology and quantifies them in valence ratings. Together with the coefficient values in the alpha frequency band associated with the motion and motivation components, the association between physiological changes in the alpha frequency band and psychological measurements from the movements cannot be a result of motor cortex activation, since Mu rhythms (8–11 Hz) are related to the activation and were only positively associated with the motivation component, and not with the motion component.

\begin{table}[t!]
    \renewcommand{\arraystretch}{1.3}
    \caption{Mean percentages of significant causation ($p < 0.05$) from the three (motion, motivation, and contentment) components to EEG spectral power in the four frequency bands (theta, alpha, beta, and gamma) for all participants.}
    \label{table:VAComponentsCausality}
    \centering
\begin{tabular}{ccccccc}
\hline 
\hline 
& \multicolumn{2}{c}{Motion} & \multicolumn{2}{c}{Motivation} & \multicolumn{2}{c}{Contentment} \\ 
\cline{2-7} 
Frequency & F3 & F4 & F3 & F4 & F3 & F4 \\
\hline 
Theta       & $\textbf{0.44}$ & $\textbf{0.38}$ & 0.39 & 0.33 & $\textbf{0.35}$ & $\textbf{0.19}$ \\
\hline
Alpha       & 0.38 & 0.31 & $\textbf{0.63}$ & $\textbf{0.48}$ & 0.21 & 0.16 \\
\hline
Beta        & 0.11 & 0.17 & 0.28  & 0.25 & 0.14 & 0.12\\
\hline
Gamma	    & 0.37 & 0.21 & 0.22 & 0.31 & 0.12 & 0.07 \\ 
\hline
\end{tabular}
\end{table}

The characteristics of the brain signals under different affective labels were investigated and analyzed by the above correlation. EEG-based statistical analysis revealed that physiological responses correlate to continuous affective labels.

\subsection{Discussion}
Our empirical study showed that the proposed A-Situ can provide affective labels underlying emotional behaviors based on visual measurement and showed the efficacy of affective curves as a reliable representation for labeling emotions. Since our system is underlain by a particular motivational theory, however, it may not cover all of the complexity or real emotion. For instance, some negative emotions such as anger cannot be measured instantly by our system, since they involve approach to (as opposed to avoidance of) negative stimuli. Some emotions related to high arousal and low movement (i.e., fear and freezing) may not be labeled as the same precision at as the SAM ratings in the valence-arousal dimensional space.

Nevertheless, our system enables people to understand how their emotions change when they feel under the situation, since the proposed labeling system outputs a set of affective labels in a spatiotemporal situation rather than a single universal set of labels. Each participant had their own emotional behaviors to recognize and deal with emotion, and such responses could be represented as continuous pairs of labels on their own affective curve by our system. The pairs of labels on an affective curve contain emotional traces in response to the affective situation, enabling our system to provide a better understanding of affective perception in a situation than existing subjective self-reports do.

Though EEG-based statistical analysis, we clarified that the affective labels contain physiological characteristics originating from psychological phenomena. In our work, EEG was the most suitable choice among available physiological measurements, which also include skin conductance, heart rate, and EMG, since it measures the brain dynamics that control thoughts, feelings, and behaviors. Using functional near-infrared spectroscopy(fNIRS) can be an alternative brain sensor to describe emotion elicitation mechanism~\cite{balconi2015hemodynamic}. However, its low temporal resolution compared to EEG has limited to measure the brain dynamics associated with emotional changes.    
\section{Conclusion}
Here, we presented a computational framework called A-Situ that provides affective labels for real-life situations, defining the term ``affective situation'' as a specific arrangement of affective entities people encounter, interact with, and which elicit some emotional response in the people. Our system showed efficacy at capturing EEG-based physiological characteristics and understanding psychological behaviors as measured by our proposed wearable device, based on real-world experiments. Modeling affective situations allows us to better understand the contents of human interactions, and representing these situations can determine the level of an interactant’s expected feelings based on the interaction. Therefore, our framework helps to bridge the semantic gap between cognitive and affective perception in real-world situations.
\section*{Acknowledgment}
This work was supported by Institute of Information \& Communications Technology Planning (IITP) grant funded by the Korea government(MSIT) (No.2017-0-00432). 

\ifCLASSOPTIONcaptionsoff
  \newpage
\fi



\bibliographystyle{IEEEtran}
\bibliography{IEEEabrv,./2018_TAC_ASITU}

\begin{IEEEbiography}[{\includegraphics[width=1in,height=1.25in,clip,keepaspectratio]{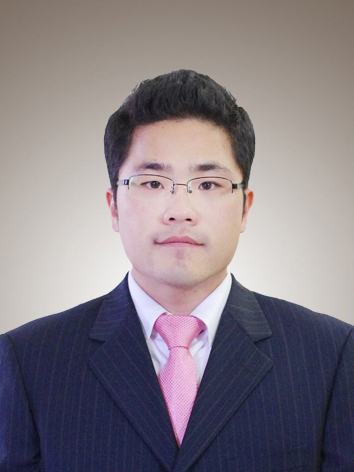}}]{Byung Hyung Kim}
received the B.S. degree in computer science from Inha University, Incheon, Korea, in 2008, and the M.S. degree in computer science from Boston University, Boston, MA, USA, in 2010. He is currently working toward the Ph.D. degree at KAIST, Daejeon, Korea. His research interests include affective computing, brain-computer interface, computer vision, assistive and rehabilitative technology, and cerebral asymmetry and the effects of emotion on brain structure.
\end{IEEEbiography}

\begin{IEEEbiography}[{\includegraphics[width=1in,height=1.25in,clip,keepaspectratio]{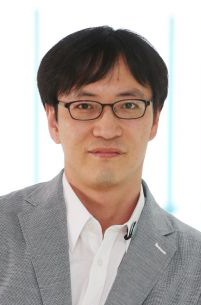}}]{Sungho Jo}
(M'09) received the B.S. degree in school of mechanical \& aerospace engineering from the Seoul National University, Seoul, Korea, in 1999, the S.M. in mechanical engineering, and Ph.D. in electrical engineering and computer science from the Massachusetts Institute of Technology (MIT), Cambridge, MA, USA, in 2001 and 2006 respectively. While pursuing the Ph.D., he was associated with the Computer Science and Artificial Intelligence Laboratory (CSAIL), Laboratory for Information Decision and Systems (LIDS), and Harvard-MIT HST NeuroEngineering Collaborative. Before joining the faculty at KAIST, he worked as a postdoctoral researcher at MIT media laboratory. Since December in 2007, he has been with the department of computer science at KAIST, where he is currently Associate Professor. His research interests include intelligent robots, neural interfacing computing, and wearable computing.
\end{IEEEbiography}

\end{document}